\documentclass[10pt,a4paper,twoside=semi,twocolumn,mpinclude=false,footinclude=false]{scrartcl}
\usepackage[left=1.5cm,right=2.2cm,top=3.4cm,bottom=3cm]{geometry}
\usepackage[mono=false]{libertine}
\usepackage[utf8]{inputenc}
\usepackage{amstext}
\usepackage{amsmath}
\usepackage{bbm}
\usepackage{theorem}
\usepackage{graphicx}
\usepackage{tikz}
\usepackage{pgfbaseplot}
\usepackage[backend=biber,style=numeric,sorting=nyt,maxnames=4]{biblatex}
\usepackage{fancyhdr}
\usepackage[noend]{algpseudocode}
\usepackage{xparse}
\usepackage{xcolor}
\usepackage{colorschemes}
\usepackage{xspace}
\usepackage{microtype}
\usepackage{booktabs}
\usepackage[many]{tcolorbox}
\usepackage{varwidth}
\usepackage[final,color,notref,notcite]{showkeys}
\usepackage{inconsolata}
\usepackage[nomessages]{fp}

\DeclareFieldFormat
  [article,inbook,incollection,inproceedings,patent,thesis,unpublished,misc]
  {title}{\mkbibquote{{\textit{#1}}}}

\colorlet{algotitlebg}{Aluminium3}
\colorlet{algotitlefg}{black}
\colorlet{algocontentbg}{Aluminium1!10!white}
\colorlet{algocontentfg}{black}
\colorlet{algoframe}{Aluminium3}
\colorlet{spotcolor}{Aluminium6}
\newcommand{\spotcolor}{\color{spotcolor}}

\addtokomafont{title}{\sffamily\spotcolor}
\addtokomafont{paragraph}{\spotcolor}
\addtokomafont{section}{\spotcolor}
\addtokomafont{subsection}{\spotcolor}
\addtokomafont{subsubsection}{\spotcolor}
\addtokomafont{descriptionlabel}{\spotcolor}
\setkomafont{caption}{}
\setkomafont{captionlabel}{}
\KOMAoptions{numbers=noenddot}

\usetikzlibrary{plotmarks}
\usetikzlibrary{arrows}
\usetikzlibrary{positioning}
\usetikzlibrary{trees}

\tikzset{task/.style={
    rectangle,
    draw=Aluminium4,
    top color=Aluminium1!25!white,
    bottom color=Aluminium1,
    font=\footnotesize,
    inner sep=2pt,
    outer sep=1pt,
    line width=.1pt}}

\colorlet{refkey}{Aluminium4}
\colorlet{labelkey}{ScarletRed3}

\renewcommand{\epsilon}{\varepsilon}
\renewcommand{\phi}{\varphi}

\NewDocumentCommand{\mc}{m}{\ensuremath{\mathcal{#1}}\xspace}
\NewDocumentCommand{\mcH}{}{\ensuremath{\mc{H}}\xspace}

\NewDocumentCommand{\mcP}{}{\ensuremath{\mc{P}}\xspace}
\NewDocumentCommand{\mcL}{}{\ensuremath{\mc{L}}\xspace}

\NewDocumentCommand{\mcI}{}{\ensuremath{\mc{I}}\xspace}
\NewDocumentCommand{\mcT}{}{\ensuremath{\mc{T}}\xspace}
\NewDocumentCommand{\mcU}{}{\ensuremath{\mc{U}}\xspace}

\NewDocumentCommand{\set}{m}{\ensuremath{\left\{ #1 \right\}}\xspace}

\NewDocumentCommand{\rank}{}{\ensuremath{\operatorname{rank}}\xspace}
\NewDocumentCommand{\parent}{m}{\ensuremath{\operatorname{parent}\left(#1\right)}\xspace}

\NewDocumentCommand{\N}{}{\ensuremath{\mathbbm{N}}}
\NewDocumentCommand{\R}{}{\ensuremath{\mathbbm{R}}}

\NewDocumentCommand{\op}{m}{\ensuremath{\operatorname{#1}}\xspace}
\NewDocumentCommand{\task}{m}{\textbf{task}(#1)\xspace}
\NewDocumentCommand{\troot}{m}{\ensuremath{\op{root}(#1)}\xspace}
\NewDocumentCommand{\sons}{m}{\ensuremath{\op{sons}(#1)}\xspace}
\NewDocumentCommand{\idep}{}{\op{in}\xspace}
\NewDocumentCommand{\odep}{}{\op{out}\xspace}

\NewDocumentCommand{\tdep}{}{\raisebox{.5ex}{\tikz\draw[line width=1pt,->,>=latex'](0,0) -- ++(1em,0);}\xspace}

\NewDocumentCommand{\function}{ m o }{\textnormal{{\ttfamily\textsc{#1}\IfNoValueF{#2}{(#2)}}}\xspace}

\newtcolorbox[auto counter]{algorithmbox}[2][]{  
  floatplacement=htb,
  float,
  enhanced,
  size=fbox,
  before skip=3mm,
  colbacktitle=algotitlebg!25!white,
  title style={top color=algotitlebg!20!white,bottom color=algotitlebg!30!white},
  colback=algocontentbg,
  colframe=algoframe,
  arc=1pt,
  halign=flush left,
  coltitle=algotitlefg,
  title={\textbf{Algorithm~\thetcbcounter:} #2},
  label={#1}}
\newtcolorbox[auto counter]{inlinealgorithmbox}{ 
  size=tight,
  boxsep=0.5mm,
  before skip=2mm,
  after skip=2mm,
  colback=algocontentbg,
  colframe=algoframe,
  arc=1pt,
  halign=flush left}
\newtcolorbox[auto counter]{figurealgorithmbox}{ 
  size=tight,
  before skip=0mm,
  colback=algocontentbg,
  boxrule=0pt,
  colframe=white,
  frame hidden,
  halign=flush left}
\NewDocumentEnvironment{inlinealgorithm}%
                       {}%
                       {\begin{inlinealgorithmbox}
                           \hspace{-0.9em}\begin{varwidth}{\linewidth}\ttfamily\small%
                           \begin{algorithmic}}%
                       {\end{algorithmic}\end{varwidth}\end{inlinealgorithmbox}}
\NewDocumentEnvironment{algorithm}%
                       {m m}%
                       {\begin{algorithmbox}[#2]{#1}%
                           \hspace{-1em}\begin{varwidth}{\linewidth}\ttfamily\small%
                           \begin{algorithmic}}%
                       {\end{algorithmic}\end{varwidth}\end{algorithmbox}}
\NewDocumentEnvironment{figurealgorithm}%
                       {}%
                       {\begin{figurealgorithmbox}
                           \hspace{-.75em}\begin{varwidth}{\linewidth}\ttfamily\small%
                           \begin{algorithmic}}%
                       {\end{algorithmic}\end{varwidth}\end{figurealgorithmbox}}

\usepgflibrary{fpu}

\NewDocumentCommand{\speedup}{m m}{%
  \pgfkeys{/pgf/fpu=true}%
  \pgfmathparse{#1/#2}%
  \pgfmathfloattofixed{\pgfmathresult}%
  \FPeval{\result}{round(\pgfmathresult,2)}%
  \result%
  \pgfkeys{/pgf/fpu=false}}

\theoremstyle{plain}
\theoremheaderfont{\bfseries \upshape}
\newtheorem{define}{Definition}[section]
\newtheorem{remark}[define]{Remark}

\setkomafont{author}{\normalsize}
\setkomafont{date}{\normalsize}

\renewenvironment{abstract}{%
  \centering
  \setlength{\parindent}{0pt}
  \begin{minipage}{\linewidth}%
    \setlength{\parindent}{0pt}%
    \setlength{\parskip}{.5em}%
    \textbf{Abstract}%
  }{\end{minipage}
}

\NewDocumentCommand{\clt}{}{\ensuremath{\tau}\xspace}
\NewDocumentCommand{\cls}{}{\ensuremath{\sigma}\xspace}
\NewDocumentCommand{\clr}{}{\ensuremath{\rho}\xspace}

\begin{document}

\title{Semi-Automatic Task Graph Construction for $\mcH$-Matrix Arithmetic}
\author{%
  \parbox{.27\textwidth}{\centering%
    {\large Steffen Börm}\small\\
    Department of Mathematics\\
    University of Kiel\\
    boerm@math.uni-kiel.de
  }
  \parbox{.33\textwidth}{\centering%
    {\large Sven Christophersen}\small\\
    Department of Mathematics\\
    University of Kiel\\
    christophersen@math.uni-kiel.de
  }
  \parbox{.27\textwidth}{\centering%
    {\large Ronald Kriemann}\small\\
    MPI for Mathematics i.t.S.\\
    Leipzig, Germany\\
    rok@mis.mpg.de
  }
}
\maketitle

\begin{abstract}
  A new method to construct task graphs for \mcH-matrix arithmetic is introduced, which uses the information
  associated with all tasks of the standard recursive \mcH-matrix algorithms, e.g., the block index set of
  the matrix blocks involved in the computation. Task refinement, i.e., the replacement of tasks by
  sub-computations, is then used to proceed in the \mcH-matrix hierarchy until the matrix blocks containing
  the actual matrix data are reached. This process is a natural extension of the classical, recursive way
  in which \mcH-matrix arithmetic is defined and thereby simplifies the efficient usage of many-core
  systems. Examples for standard and accumulator based \mcH-arithmetic are shown for model problems with
  different block structures.

  \textbf{AMS Subject Classification:} 65F05, 65Y05, 65Y20, 68W10, 68W40 \\
  \textbf{Keywords:} hierarchical matrices, task graph, parallel algorithms, many-core processors

\end{abstract}

\pagestyle{fancy}
\thispagestyle{plain}
\fancyhf{}
\fancyhf[HLE]{\footnotesize\thepage\hfill S. Börm, S. Christophersen and R. Kriemann}
\fancyhf[HRO]{\footnotesize DAG Generation for \mcH-matrix Arithmetic \hfill\thepage}
\fancyhf[FC]{}
\renewcommand{\headrulewidth}{0.4pt}
\renewcommand{\footrulewidth}{0pt}
\thispagestyle{empty}



\section{Introduction} \label{sec:intro}

Hierarchical matrices (\mcH-matrices), introduced in \cite{Hackbusch:1999}, are a powerful tool to represent dense
matrices coming from integral equations or partial differential equations in a hierarchical, block-oriented, data-sparse
way with log-linear memory costs. Furthermore, a matrix arithmetic, e.g., matrix addition, multiplication, inversion and
factorization, is possible with log-linear computation costs (see \cite{GrasedyckHackbusch:2003}).

Classical arithmetic for \mcH-matrices is formulated recursively following the recursive block structure of the
matrices. This formulation has the advantage of simplicity, since only local blocks are addressed, e.g., the sub blocks
of the current matrix block, and therefore the implementation only needs to handle a few of them. The latter also
simplifies the analysis of the arithmetic and their implementation.

In \cite{Bor:2019} a modified formulation of the \mcH-arithmetic was introduced, which collects all updates to sub
blocks in \emph{accumulators}, thereby postponing the modification of those sub blocks only after all updates are
available. Furthermore, the application of these accumulated updates strictly follows the hierarchy of the \mcH-matrix,
pushing updates to structured matrix blocks only to the next level below. With this, the number of updates applied to
leaf blocks of the \mcH-matrix is reduced and such also the number of low-rank truncations. This significantly
improves the runtimes of \mcH-arithmetic.

Due to the substantial changes in the hardware landscape in the last decade, e.g., with many-core CPUs integrating 64
and more cores into a single CPU, e.g., AMD Epyc 7002 series, the implementation of \mcH-matrix arithmetic also
needs to efficiently make use of thread-level parallelism to speed up the \mcH-matrix computations. However, using the
recursive functions and applying parallelization on the local level as used in \cite{Kriemann:2005} introduces too much
artificial synchronisation points to be efficient with such a high number of CPU cores.

Therefore, a different strategy is used for many-core CPUs based on \emph{tasks} to describe the atomic computation
blocks and their dependencies which form a directed acyclic graph (DAG). This task graph is handed to a scheduling system
to execute a task when all its dependencies are met on the next free CPU core. Such task-based approaches were also used
for dense \cite{Buttari:2008,Buttari:2009} and sparse \cite{Hogg:2010,Lac:2012,But:2013} arithmetic and previously
described in \cite{Kri:2013} for \mcH-matrices.

\begin{figure}
  \centering
  \begin{tikzpicture}[xscale=0.35,yscale=-0.35]

    \fill[ScarletRed1] (0.5,0.5) rectangle ++(0.5,0.5);

    \fill[SkyBlue1] (5.5,0.5) rectangle ++(0.5,0.5); 
    \fill[SkyBlue1] (4.0,0) rectangle ++(1,1); 
    \fill[SkyBlue1] (0,4.0) rectangle ++(2,2); 

    \fill[Chameleon1] (4,4) rectangle ++(2,2);
    \fill[Chameleon3] (4,4) rectangle ++(1,1);
    \fill[Chameleon3] (5.5,5.5) rectangle ++(0.5,0.5);
    
    \draw (0,0) rectangle ++(8,8);
    \foreach \x in {0,2,4,6} {
      \foreach \y in {0,2,4,6} {
        \draw (\x,\y) rectangle ++(2,2);
      }
    }
    \foreach \x in {0,1,...,7} {
      \draw (\x,\x) rectangle ++(1,1);
    }
    \draw (7,6) rectangle ++(1,1); 

    \draw (0,0) rectangle ++(0.5,0.5);
    \draw (0.5,0) rectangle ++(0.5,0.5);
    \draw (0,0.5) rectangle ++(0.5,0.5);
    \draw (0.5,0.5) rectangle ++(0.5,0.5);

    \draw (4,0) rectangle ++(1,1);
    \draw (5,0) rectangle ++(1,1);
    \draw (4,1) rectangle ++(1,1);
    \draw (5,0) rectangle ++(0.5,0.5);
    \draw (5.5,0) rectangle ++(0.5,0.5);
    \draw (5,0.5) rectangle ++(0.5,0.5);
    \draw (5.5,0.5) rectangle ++(0.5,0.5);

    \foreach \x in {4,4.5,5,5.5} {
      \draw (\x,\x) rectangle ++(0.5,0.5);
    }


  \end{tikzpicture}
  \quad
  \begin{tikzpicture}[xshift=5cm,scale=0.8,
    every node/.style={fill=black,rectangle,inner sep=1.25pt,outer sep=0pt},
    level 1/.style={sibling distance=1.75cm, level distance=8mm},
    level 2/.style={sibling distance=3mm, level distance=8mm},
    level 3/.style={sibling distance=3mm, level distance=8mm}]
    level 4/.style={sibling distance=3mm, level distance=8mm}]
    \node {}
    child [sibling distance=1.8cm] { node {}
      child { node {} 
        child [sibling distance=2.5mm] { node {} 
          child [sibling distance=2mm] { node {} 
          }
          child [sibling distance=2mm] { node {}
          }
          child [sibling distance=2mm] { node {}
          }
          child [sibling distance=2.5mm] { node (lu) [draw,fill=ScarletRed1,inner sep=2.5pt] {}
          }
        }
        child [sibling distance=2mm] { node {}
        }
        child [sibling distance=2mm] { node {}
        }
        child [sibling distance=2mm] { node {}
        }
      }
      child { node {}
      }
      child [sibling distance=3mm] { node {}
      }
      child [sibling distance=3mm] { node {}
        child [sibling distance=2mm] { node {} 
        }
        child [sibling distance=2mm] { node {}
        }
        child [sibling distance=2mm] { node {}
        }
        child [sibling distance=2mm] { node {}
        }
      }
    }
    child [sibling distance=1.25cm] { node {}
      child [sibling distance=3mm] { node {}
        child { node (solve1) [draw,fill=SkyBlue1,inner sep=2.5pt] {} 
        }
        child { node {}
          child [sibling distance=2mm] { node {} 
          }
          child [sibling distance=2mm] { node {}
          }
          child { node {}
          }
          child { node (solve2) [draw,fill=SkyBlue1,inner sep=2.5pt] {}
          }
        }
        child [sibling distance=2mm] { node {}
        }
        child [sibling distance=2mm] { node {}
        }
      }
      child [sibling distance=3mm] { node {}
      }
      child [sibling distance=2mm] { node {}
      }
      child [sibling distance=2mm] { node {}
      }
    }
    child [sibling distance=0.8cm] { node {}
      child [sibling distance=2.5mm] { node (solve3) [draw,fill=SkyBlue1,inner sep=2.5pt] {}
      }
      child [sibling distance=2mm] { node {}
      }
      child [sibling distance=2mm] { node {}
      }
      child [sibling distance=2mm] { node {}
      }
    }
    child [sibling distance=1.0cm] { node {}
      child { node (update1) [draw,fill=Chameleon1,inner sep=2.5pt] {}
        child [sibling distance=3mm] { node (update2) [draw,fill=Chameleon3,inner sep=2.5pt] {} 
          child [sibling distance=2mm] { node {} 
          }
          child [sibling distance=2mm] { node {}
          }
          child [sibling distance=2mm] { node {}
          }
          child [sibling distance=2mm] { node {}
          }
        }
        child [sibling distance=2mm] { node {}
        }
        child [sibling distance=2mm] { node {}
        }
        child [sibling distance=2.5mm] { node {}
          child [sibling distance=2mm] { node {} 
          }
          child [sibling distance=2mm] { node {}
          }
          child [sibling distance=2mm] { node {}
          }
          child [sibling distance=2.5mm] { node (update3) [draw,fill=Chameleon3,inner sep=2.5pt] {}
          }
        }
      }
      child { node {}
      }
      child { node {}
      }
      child { node {}
        child [sibling distance=2mm] { node {} 
        }
        child [sibling distance=2mm] { node {}
        }
        child [sibling distance=2mm] { node {}
        }
        child [sibling distance=2mm] { node {}
        }
      }
    };

    \begin{scope}[line width=.5pt,>=latex',->]
      \draw [ScarletRed2] (lu) -- (solve1);
      \draw [ScarletRed2] (lu) .. controls (-1.5,-3.6) .. (solve2);
      \draw [ScarletRed2] (lu) .. controls (-0.5,-3.9) and (0,-3.6) .. (solve3);

      \draw [SkyBlue2] (solve1) .. controls (-1,-2.75) and (0,-2.75) .. (update2);
      \draw [SkyBlue2] (solve2) .. controls (0,-3.6) and (1,-3.6) .. (update3);
      \draw [SkyBlue2] (solve3) .. controls (0.3,-1.8) and (0.7,-1.8) .. (update1);
    \end{scope}
  \end{tikzpicture}
  \caption{\mcH-matrix (left) and dependencies of matrix blocks during \mcH-LU in tree structure (right).}
  \label{fig:hludep}
\end{figure}
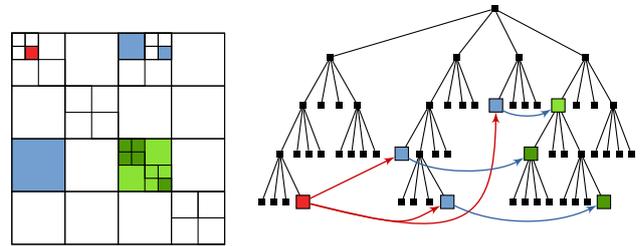

The remaining problem is to construct the DAG for the runtime scheduling system. Constructing the DAG includes
identification of the compute tasks and especially their dependencies. Normally, both are based on the arithmetical
formulation of the data the tasks work on. For many dense or sparse matrix algorithms without a complex recursive
hierarchy, the dependencies can often be directly expressed, e.g., based on matrix blocks or coefficient indices. For
\mcH-matrix arithmetic, it is more complicated because the matrix blocks are defined on different levels of the
hierarchy. An example is shown in Figure~\ref{fig:hludep}. There, for the \mcH-LU factorization, the red diagonal block
forms a dependency for the blue blocks, which for themselves form a dependency for the updates of the green matrix
blocks. In the corresponding tree representing the relation between all matrix blocks, the blocks
are on different levels and not necessarily close to each other. The connecting paths may go back to the root of the
tree. Furthermore, inner blocks of the tree do not correspond to actual data, as this is stored only at the leaf blocks,
and hence the computation affects all sub-blocks, thereby creating more dependencies.

This is the reason why in \cite{Kri:2013}, the traditional formulation of \mcH-matrix arithmetic was changed
to have a \emph{level-wise}, global view of the matrix similar to single level dense arithmetic, i.e., on each level of
the hierarchy, all matrix blocks in a block row or column were used to set up dependencies. The resulting task graph
represented data dependency over the whole \mcH-matrix and permitted to schedule ready tasks independent of the position
in the matrix without unnecessary task synchronisation. However, the modified \mcH-matrix arithmetic formulation requires
extra data to permit access to all needed matrix blocks and the process of defining the task graph was error-prone,
which hinders the implementation of DAGs for new arithmetic functions.

A more natural way of defining the DAG would be to follow the standard, recursive \mcH-matrix functions. However, this
would require to handle nested task parallelism with dependencies over different recursion paths. Various task runtime
scheduling systems exist which try to address this problem. The most widely used of such systems is OpenMP
\cite{dagum1998openmp}, which introduced tasks in v3 and extended this by task dependencies in v4 \cite{openmp4},
where data input/output dependencies are defined by memory ranges (memory address plus length). Though this works well
for single-level algorithms, dependencies between sub-tasks in different recursion paths are difficult to
implement\footnote{Fixed hierarchies would permit predefined, static task graphs. Unrestricted hierarchies requiring
  dynamic task graphs are basically impossible to implement.}

The same limitations apply to the OmpSs \cite{OmpSs} parallelization framework, which introduced the task system before
OpenMP. Despite these restrictions, OmpSs was used in \cite{AliCarKriQui:2017} to construct task graphs for \mcH-matrix
arithmetic. However, only a very restricted, non-efficient version of \mcH-matrix arithmetic was possible.

In \cite{PerBelLabAyg:2017} an extension to OpenMP, implemented in OmpSs-2 \cite{OmpSs2}, was introduced, which
distinguishes between standard and \emph{weak} dependencies. A weak dependency from a parent task to a sub-task
does not require the parent task to wait for the sub-task to finish as would be needed in OmpSs (or OpenMP) thereby
avoiding unnecessary task synchronisation. This extension would permit the implemention of nested functions with fine
grained dependencies as the \mcH-matrix arithmetic makes use of and was used in \cite{RocChrAliBelBoeQui:2019} to fully
implement task-based \mcH-matrix arithmetic. The presented numerical results demonstrate that the technique has some
potential but needs further optimizations to be efficient for a wide range of \mcH-matrix structures. Furthermore, a
special compiler is needed supporting these non-standard features, though it is expected that weak dependencies will
eventually be introduced also in OpenMP.

With so called \textit{bubbles of tasks}, StarPU (\cite{Augonnet:2011,Thibault:2018}) tries to address the issue, where
tasks are not restricted to wait for sub-tasks to finish but where these sub-tasks may extend data dependencies over
the local boundaries as defined by recursion. It is currently unclear, whether this concept is capable of efficiently
handling recursive \mcH-matrix arithmetic.



Because of these difficulties, we avoid such a general approach and propose a simpler method for \mcH-matrix functions,
which makes use of data that is coupled with all sub-blocks of \mcH-matrices: the block index sets. With the block index
sets for all input and output matrices of a function the problem of addressing the actual data storage vanishes as any
dependency automatically includes any (leaf) sub-block. Furthermore, corresponding data dependencies are automatically
constructed and refined when replacing tasks by sub-tasks, e.g., during function recursion. During the refinement, the
dependencies can be filtered based on sub-set tests for the block index sets of the sub-blocks associated with the
sub-tasks. This eliminates unneeded dependencies to parent or sibling tasks. As a result, a task graph for \mcH-matrix
arithmetic is computed which can spawn tasks for leaf blocks as soon as possible and avoids unneccessary
synchronization.

Furthermore, since the task graph is constructed without a particular task scheduling system, the new method can be
combined with an arbitrary task runtime system. Therefore, for a particular computer system, the best runtime system may
be chosen.

This article is structured as follows: in Section~\ref{sec:hmat} \mcH-matrices and their arithmetic are
introduced. The new DAG construction is described in Section~\ref{sec:tasks} with some optimizations presented in
Section~\ref{sec:opttech}. Section~\ref{sec:numexp} contains the results of several numerical experiments comparing
the different approaches.

All presented algorithms are available in the software HLR (see \cite{HLR}) released under an open-source license.


\section{\mcH-Matrices and \mcH-Arithmetic} \label{sec:hmat}


\subsection{Definitions} \label{sec:hdef}

For an indexset \(I\) we define the \emph{cluster tree} (or \mcH-tree) as the hierarchical partitioning of \(I\) into
disjoint sub-sets of \(I\):
\begin{define}[Cluster Tree]
  Let \(T_I = (V,E)\) be a tree with \(V \subset \mathcal{P}(I)\). \(T_I\) is called a \emph{cluster tree} over \(I\) if
  \begin{enumerate}
  \item \(I = \troot{T_I}\) and
  \item for all \(v \in V\) with \(\sons{v} \ne \emptyset : v = \dot\cup_{v' \in \sons{v}} v'\).
  \end{enumerate}
  A node in \(T_I\) is also called a \emph{cluster} and we write \(\clt \in T_I\) if \(\clt \in V\). The set of leaves
  of \(T_I\) is denoted by \(\mcL(T_I)\). 
\end{define}

Similar to a cluster tree we can extend the hierarchical partitioning to the product \(I \times J\) of two index sets
\(I, J\), while restricting the possible set of nodes by given cluster trees \(T_I\) and \(T_J\) over \(I\) and \(J\),
respectively. Furthermore, the set of leaves will be defined by an application dependent \emph{admissibility condition}
(see \cite{Hackbusch:2015} for examples).

\begin{define}[Block Tree]
  Let \(T_I, T_J\) be two cluster trees and let \(\op{adm} : T_I \times T_J \to \mathbbm{B} \). The \emph{block tree}
  \(T = T_{I \times J} \) is recursively defined starting with \( \troot{T} = (I,J) \):
  \begin{align*}
    & \sons{\clt,\cls} = \\
    & \qquad \begin{cases}
      \emptyset \textnormal{ if } \op{adm}(\clt,\cls) = \textnormal{true}, \sons{\clt} = \emptyset \textnormal{ or } \sons{\cls} = \emptyset, \\
      \set{ (\clt',\cls') \,:\, \clt' \in \sons{\clt}, \cls' \in \sons{\cls} } \textnormal{ else}.
    \end{cases}
  \end{align*}
  A node in \(T\) is also called a \emph{block}. Again, the set of leaves of \(T\) is denoted by \(\mcL(T) := \set{ b
    \in T \;:\; \sons{b} = \emptyset}\).
\end{define}

The admissibility condition ensures that admissible blocks in $T$, i.e., blocks \(b\) with \(\op{adm}(b) =
\textnormal{true}\), can be approximated by a predefined rank \(k\) (or up to a predefined accuracy
\(\varepsilon\)). The set of all such matrices forms the set of \mcH-matrices:
\begin{define}[\mcH-Matrix]
  For a block tree $T$ over cluster trees $T_I, T_J$ and $k \in \N$, the set of \mcH-matrices $\mcH(T,k)$ is defined as
  \begin{align*}
    \mcH(T,k) := & \left\{ M \in \R^{I \times J} \;:\; \forall (\clt,\cls) \in \mcL(T) : \right. \\
    & \left. \rank(M_{\clt,\cls}) \le k \vee \clt \in \mcL(T_I) \vee \cls \in \mcL(T_J) \right\}
  \end{align*}
  Here, \(M_{\clt,\cls}\) refers to the sub-block \(M|_{\clt \times \cls}\).
\end{define}


\subsection{\mcH-Arithmetic}

For many arithmetical functions the matrix multiplication forms the basic building block. In this work, we will
consider the general version
\begin{displaymath}
  C := \alpha A \cdot B + C
\end{displaymath}
which applies the update \(\alpha A B\) to the matrix \(C\). If not stated otherwise, we will assume a binary cluster
tree, e.g., for a non-leaf cluster \(t\) we have \(\sons{\clt} = \set{\clt_0,\clt_1}\), and hence a quad block cluster tree,
which will simplify the presentation. The algorithms can easily be extended for general cluster trees.

For an \mcH-matrix \(M_{\clt,\cls} \in \mcH(T)\) with \(T\) based on a binary tree, the block structure can be written as
\begin{displaymath}
  M =
  \begin{pmatrix}
    M_{\clt_0,\cls_0} & M_{\clt_0,\cls_1} \\
    M_{\clt_1,\cls_0} & M_{\clt_1,\cls_1}
  \end{pmatrix}
\end{displaymath}

Using this notation for the above matrix multiplication, the algorithm for the \mcH-matrix multiplication can
be written recursively as

\begin{algorithm}{\mcH-Matrix Multiplication}{alg:hmul}
  \Procedure{hmul}{in: $\alpha,A_{\clt,\rho}, B_{\rho,\cls}$, inout: $C_{\clt,\cls}$}
  \If{\(\set{(\clt,\rho),(\rho,\cls),(\clt,\cls)} \cap \mcL(T) = \emptyset \)}
  \For{\(i,j,\ell \in \set{0,1}\)}
  \State \function{hmul}[\(\alpha,A_{\clt_i,\rho_\ell},B_{\rho_\ell,\cls_j},C_{\clt_i,\cls_j}\)];
  \EndFor
  \Else
  \State \(C_{\clt,\cls} := C_{\clt,\cls} + \alpha A_{\clt,\rho} B_{\rho,\cls}\);
  \EndIf
  \EndProcedure
\end{algorithm}

In the non-recursive part, special routines will handle the different multiplications between structured,
dense and low-rank matrices.

An only slightly more advanced matrix algorithm is the LU factorization \(A_{\clt,\clt} = L_{\clt,\clt} U_{\clt,\clt}\)
of the matrix \(A_{\clt,\clt}\) into triangular factors \(L_{\clt,\clt}\) and \(U_{\clt,\clt}\). Using the above block
structure for the \mcH-matrix \(A_{\clt,\clt}\), this reads
\begin{displaymath}\small
  \begin{pmatrix}
    A_{\clt_0,\clt_0} & A_{\clt_0,\clt_1} \\
    A_{\clt_1,\clt_0} & A_{\clt_1,\clt_1}
  \end{pmatrix}
  =
  \begin{pmatrix}
    L_{\clt_0,\clt_0} &  \\
    L_{\clt_1,\clt_0} & U_{\clt_1,\clt_1}
  \end{pmatrix}
  \begin{pmatrix}
    U_{\clt_0,\clt_0} & U_{\clt_0,\clt_1} \\
    & U_{\clt_1,\clt_1}
  \end{pmatrix}
\end{displaymath}
which leads to Algorithm~\ref{alg:hlu} with recursive call in case of structured matrices, using functions
\function{htrsl} and \function{htrsu} for the matrix solve operations, and a dense LU factorization if the input matrix
is dense.

\begin{algorithm}{\mcH-LU factorization}{alg:hlu}
  \Procedure{hlu}{in: $A_{\clt,\clt}$, out: $L_{\clt,\clt}, U_{\clt,\clt}$}
    \If{\( (\clt,\clt) \not\in \mcL(T)\)}
      \State \function{hlu}[\(A_{\clt_0,\clt_0}, L_{\clt_0,\clt_0}, U_{\clt_0,\clt_0}\)];
      \State \function{htrsu}[\(U_{\clt_0,\clt_0}, A_{\clt_1 ,\clt_0}, L_{\clt_1 ,\clt_0}\)];
      \State \function{htrsl}[\(L_{\clt_0,\clt_0}, A_{\clt_0,\clt_1}, U_{\clt_0,\clt_1}\)];
      \State \function{hmul}[\(-1, L_{\clt_1,\clt_0}, U_{\clt_0,\clt_1}, A_{\clt_1,\clt_1}\)];
      \State \function{hlu}[\(A_{\clt_1,\clt_1}, L_{\clt_1,\clt_1}, U_{\clt_1,\clt_1}\)];
    \Else
      \State solve \(A_{\clt,\clt} = L_{\clt,\clt} U_{\clt,\clt}\);
    \EndIf
  \EndProcedure
\end{algorithm}


Coming back to the matrix solves, \(L_{\clt,\clt} X_{\clt,\cls} = M_{\clt,\cls}\) with a lower triangular matrix
\(L_{\clt,\clt}\) can be written using the block structure as
\begin{displaymath}\small
  \begin{pmatrix}
    L_{\clt_0,\clt_0} & \\
    L_{\clt_0,\clt_1} & L_{\clt_1,\clt_1}
  \end{pmatrix}
  \begin{pmatrix}
    X_{\clt_0,\cls_0} & X_{\clt_0,\cls_1} \\
    X_{\clt_0,\cls_1} & X_{\clt_1,\cls_1}
  \end{pmatrix}
  =
  \begin{pmatrix}
    M_{\clt_0,\cls_0} & M_{\clt_0,\cls_1} \\
    M_{\clt_0,\cls_1} & M_{\clt_1,\cls_1}
  \end{pmatrix}
\end{displaymath}
With \(M\) being given and \(X\) sought, we obtain the equations for the sub-blocks which can be used to
formulate the algorithm for \function{htrsl} as shown in Algorithm~\ref{alg:htrsm}.

\begin{algorithm}{Lower triangular \mcH-matrix solve}{alg:htrsm}
  \Procedure{htrsl}{in: $L_{\clt,\clt}, M_{\clt,\cls}$, out: $X_{\clt,\cls}$}
    \If{\((\clt,\cls) \not\in \mcL(T)\)}
    \State\function{htrsl}[\(L_{\clt_0,\clt_0}, M_{\clt_0,\cls_0}, X_{\clt_0,\cls_0}\)];
    \State\function{htrsl}[\(L_{\clt_0,\clt_0}, M_{\clt_0,\cls_1}, X_{\clt_0,\cls_1}\)];
    \State\function{hmul}[\(-1, L_{\clt_1,\clt_0}, X_{\clt_0,\cls_0}, M_{\clt_1,\cls_0}\)];
    \State\function{hmul}[\(-1, L_{\clt_1,\clt_0}, X_{\clt_0,\cls_1}, M_{\clt_1,\cls_1}\)];
    \State\function{htrsl}[\(L_{\clt_1,\clt_1}, M_{\clt_1,\cls_0}, X_{\clt_1,\cls_0}\)];
    \State\function{htrsl}[\(L_{\clt_1,\clt_1}, M_{\clt_1,\cls_1}, X_{\clt_1,\cls_1}\)];
    \Else
    \State solve \(L_{\clt,\clt} X_{\clt,\cls} = M_{\clt,\cls}\);
    \EndIf
  \EndProcedure
\end{algorithm}

Similarly, the function \function{htrsu} for solving \(X_{\cls,\clt} U_{\clt,\clt} = M_{\cls,\clt}\)with an upper
triangular matrix block \(U_{\clt,\clt}\) can be implemented.


\subsection{Accumulator based Arithmetic} \label{sec:accu}

In the formulation of \function{hmul} each update in the non-recursive part is applied to the destination matrix
\(C_{\clt,\cls}\) as soon as possible in standard implementations of \mcH-matrix arithmetic. For low-rank matrices
\(C_{\clt,\cls}\), each of these updates involve a truncation operation to reduce the rank of the sum \(C_{\clt,\cls} + \alpha
A_{\clt,\clr} B_{\clr,\cls}\) to the predefined rank \(k\) or precision \(\varepsilon\).

Such updates to low-rank matrices may also occur if \(C\) is a structured matrix and \(\alpha A_{\clt,\clr} B_{\clr,\cls}\) is a
low-rank update, e.g., if either \(A_{\clt,\clr}\) or \(B_{\clr,\cls}\) corresponds to a low-rank matrix. In this case, all leaf
sub-blocks of \(C_{\clt,\cls}\) will be updated. Again, each of those updates is applied individually in typical
implementation for \mcH-matrix arithmetic. This often leads to a significant number of truncation operations for
low-rank blocks within an \mcH-matrix.

In \cite{Bor:2019}, a different approach was described, where updates are collected level-wise in a separate
matrix, called \emph{accumulator}. After all updates per level are applied, these collected updates are
shifted down to the accumulators of the matrix blocks of the next level. The process is then repeated until
the leaf blocks in the matrix are reached. At this point all updates to the destination block have been
collected in the corresponding accumulator matrix and are applied in a single update step.

By collecting updates per level, the number of truncation operations can be reduced significantly. Since these
contribute to a large part of the overall runtime of typical \mcH-arithmetic functions, this also leads to
faster algorithms.

\begin{remark}
  A related modification of the \mcH-arithmetic was introduced in \cite{DoeHarMul:2019} where updates are also postponed
  until the leaf matrix blocks need to be modified. In contrast to the accumulator based \mcH-arithmetic, the updates in
  \cite{DoeHarMul:2019} are not accumulated per level of the block tree but all updates are shifted to the leaves.
\end{remark}

For \(C_{\clt,\cls}\) the accumulator matrix shall be denoted by \(\mcU_{\clt,\cls}\). \(\mcU_{\clt,\cls}\) will contain the sum of all updates
to \(C_{\clt,\cls}\) for which \(\alpha A_{\clt,\clr} B_{\clr,\cls}\) results in a low-rank or dense matrix and the update can be applied
directly. If \(\alpha A_{\clt,\clr} B_{\clr,\cls}\) results in a structured matrix, the application will be deferred to
sub-blocks of \(C_{\clt,\cls}\), which corresponds to the recursive step of Algorithm~\ref{alg:hmul}. Such updates will be
stored in the set \(\mcP_{\clt,\cls}\) of \emph{pending} updates.

The storage format of \(\mcU_{\clt,\cls}\) is left open. By default, a low-rank representation in factorised form is used, where
\(\mcU_{\clt,\cls}\) will not need storage space at the start of the arithmetic because \(\rank(\mcU_{\clt,\cls})=0\). However, for
optimisation reasons, a dense storage format may be more efficient if dense updates to \(C_{\clt,\cls}\) occur.

For the accumulator arithmetic, the handling of updates of the form \(C_{\clt,\cls} := C_{\clt,\cls} + \alpha A_{\clt,\clr} B_{\clr,\cls}\) is
split into two steps, represented by different functions. The first step is implemented by \function{add\_upd}, which
collects the update \(\alpha A_{\clt,\clr} B_{\clr,\cls}\) and either applies it to the accumulator \(\mcU_{\clt,\cls}\) if the product can
be evaluated or stores the tuple \((\alpha,A_{\clt,\clr}, B_{\clr,\cls})\) in the set \(\mcP_{\clt,\cls}\) of pending updates.

\begin{algorithm}{Collect single update}{alg:addprod}
  \Procedure{add\_upd}{\textbf{in}: $\alpha, A_{\clt,\clr}$, $B_{\clr,\cls}, C_{\clt,\cls}$}
    \If{\(\set{(\clt,\clr),(\clr,\cls),(\clt,\cls)} \cap \mcL(T) = \emptyset \)}
      \State \( \mcP_{\clt,\cls} := \mcP_{\clt,\cls} \cup \set{(\alpha,A_{\clt,\clr},B_{\clr,\cls})} \);
    \Else
      \State \( \mcU_{\clt,\cls} := \mcU_{\clt,\cls} + \alpha \cdot A_{\clt,\clr} \cdot B_{\clr,\cls} \);
    \EndIf
  \EndProcedure
\end{algorithm}

The second step consists of shifting down the collected updates in \(\mcU_{\clt,\cls}\) and \(\mcP_{\clt,\cls}\) to sub-blocks in case of
structured matrices or applying the accumulated updates to the leaf matrix \(C_{\clt,\cls}\), and is shown in
Algorithm~\ref{alg:applyupd} in function \function{apply\_upd}. The actual update shift is implemented in
Algorithm~\ref{alg:shiftupd}. There, for pending updates the individual update factors are split, corresponding to the
triple-loop in Algorithm~\ref{alg:hmul}.

\begin{algorithm}{Apply all collected updates}{alg:applyupd}
  \Procedure{apply\_upd}{\textbf{in}: $C_{\clt,\cls}$}
    \If{ \( (\clt,\cls) \not\in \mcL(T) \) }
      \State \function{shift\_upd}[\( C_{\clt,\cls} \)];
      \For{ \( \clt' \in \sons{\clt}, \cls' \in \sons{\cls} \) }
        \State \function{apply\_upd}[\(C_{\clt',\cls'}\)];
      \EndFor
    \Else
      \State \( C_{\clt,\cls} := C_{\clt,\cls} + \mcU_{\clt,\cls} \);
    \EndIf
  \EndProcedure
\end{algorithm}

\begin{algorithm}{Shift accumulated updates to sub-blocks}{alg:shiftupd}
  \Procedure{shift\_upd}{\textbf{in}: $C_{\clt,\cls}$}
    \For{ \( \clt' \in \sons{\clt}, \cls' \in \sons{\cls} \) }
      \State \( \mcU_{\clt',\cls'} := \mcU_{\clt',\cls'} + \mcU_{\clt,\cls}|_{\clt',\cls'} \) ;
      \For{ \( (\alpha,A_{\clt,\clr},B_{\clr,\cls}) \in \mcP_{\clt,\cls}, \clr' \in \sons{\clr} \) }
        \State \function{add\_upd}[\(\alpha, A_{\clt',\clr'}, B_{\clr',\cls'}, C_{\clt',\cls'}\)];
      \EndFor
    \EndFor
  \EndProcedure
\end{algorithm}

With these functions, the standard \mcH-matrix multiplication \(C := C + A B\) is evaluated by replacing the
function call
\begin{inlinealgorithm}
  \State \function{hmul}[\(1,A,B,C\)];
\end{inlinealgorithm}
{\noindent by}
\begin{inlinealgorithm}
  \State \function{add\_upd}[\(1,A,B,C\)];
  \State \function{apply\_upd}[\(C\)];
\end{inlinealgorithm}

For the \mcH-LU factorization, one could follow the same scheme and replace the function \function{hmul} by
the corresponding functions \function{add\_upd} and \function{apply\_upd}. However, this might fail to
collect all updates before applying the accumulator to the destination matrix block. The reason is, that on a
single level in the \mcH-LU factorization, multiple \function{hmul} calls may occur to the same destination,
e.g., if the block structure is not only \(2 \times 2\). Also, updates from different recursion levels of the
LU factorization are not handled.

Instead, collection and application of updates are split during \mcH-LU. Each call to \function{hmul} will be replaced
by \function{add\_upd}, e.g., only collecting the updates. If a recursive step occurs during \mcH-LU, the accumulated
updates are shifted down to all sub-blocks with \function{shift\_upd}, thereby ensuring that all sub-blocks will have
all collected updates from the upper levels. For leaf matrix blocks, the updates are applied before (dense)
factorization using \function{apply\_upd}.
  
\begin{algorithm}{\mcH-LU factorization with accumulators}{alg:hluaccu}
  \Procedure{hlu}{in: $A_{\clt,\clt}$, out: $L_{\clt,\clt}, U_{\clt,\clt}$}
    \If{\( (\clt,\clt) \not\in \mcL(T)\)}
      \State \function{shift\_upd}[\(A_{\clt,\clt}\)];
      \State \function{hlu}[\(A_{\clt_0,\clt_0}, L_{\clt_0,\clt_0}, U_{\clt_0,\clt_0}\)];
      \State \function{htrsu}[\(U_{\clt_0,\clt_0}, A_{\clt_1,\clt_0}, L_{\clt_1,\clt_0}\)];
      \State \function{htrsl}[\(L_{\clt_0,\clt_0}, A_{\clt_0,\clt_1}, U_{\clt_0,\clt_1}\)];
      \State \function{add\_upd}[\(-1, L_{\clt_1,\clt_1}, U_{\clt_1,\clt_1}, A_{\clt_1,\clt_1} \)];
      \State \function{hlu}[\(A_{\clt_1,\clt_1}, L_{\clt_1,\clt_1}, U_{\clt_1,\clt_1}\)];
    \Else
      \State \function{apply\_upd}[\(A_{\clt,\clt}\)];
      \State \(A_{\clt,\clt} = L_{\clt,\clt} U_{\clt,\clt}\);
    \EndIf
  \EndProcedure
\end{algorithm}

The same strategy is applied for the matrix solve functions, e.g., only collect updates whenever
\function{hmul} is called and shift (apply) updates at each recursive (non-recursive) step.

\begin{algorithm}{Lower Triangular \mcH-Matrix Solve with Accumulators}{alg:htrsmaccu}
  \Procedure{htrsl}{in: $L_{\clt,\clt}, M_{\clt,\cls}$, out: $X_{\clt,\cls}$}
    \If{\((\clt,\cls) \not\in \mcL(T)\)}
      \State\function{shift\_upd}[\(M_{\clt,\cls}\)];
      \State\function{htrsl}[\(L_{\clt_0,\clt_0}, M_{\clt_0,\cls_0}, X_{\clt_0,\cls_0}\)];
      \State\function{htrsl}[\(L_{\clt_0,\clt_0}, M_{\clt_0,\cls_1}, X_{\clt_0,\cls_1}\)];
      \State\function{add\_upd}[\(-1, L_{\clt_1,\clt_0}, X_{\clt_0,\cls_0}, M_{\clt_1,\cls_0}\)];
      \State\function{add\_upd}[\(-1, L_{\clt_1,\cls_0}, X_{\clt_0,\cls_1}, M_{\clt_1,\cls_1}\)];
      \State\function{htrsl}[\(L_{\clt_1,\clt_1}, M_{\clt_1,\cls_0}, X_{\clt_1,\cls_0}\)];
      \State\function{htrsl}[\(L_{\clt_1,\clt_1}, M_{\clt_1,\cls_1}, X_{\clt_1,\cls_1}\)];
    \Else
      \State\function{apply\_upd}[\(M_{\clt,\cls}\)];
      \State solve \(L_{\clt,\clt} X_{\clt,\cls} = M_{\clt,\clt}\);
    \EndIf
  \EndProcedure
\end{algorithm}


\section{Task based \mcH-Arithmetic} \label{sec:tasks}

\subsection{Task refinement} \label{sec:taskrefine}

For all \mcH-matrix arithmetic functions \(f\), e.g., \function{hmul} or \function{hlu}, we can define a corresponding
task \task{\(f\)}. For simplicity, we will subsequently identify the \mcH-arithmetic function with its task, e.g., write
\function{hlu} instead of \task{\function{hlu}}, if no ambiguity between both concepts exists.

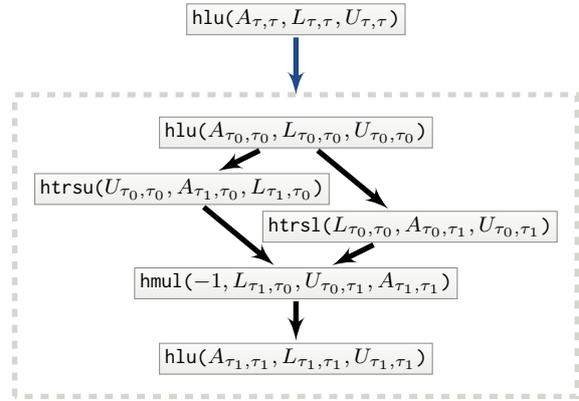
\begin{figure}
  \centering
  \begin{tikzpicture}[line width=1pt,>=latex',every node/.style=task]

    \begin{scope}[line width=2pt]
      \node (lu0)    [anchor=center] at (0,0)   {\function{hlu}[\(A_{\clt,\clt},L_{\clt,\clt},U_{\clt,\clt}\)]};
      \draw [SkyBlue3,->] (lu0) -- ++(0,-1.0);
    \end{scope}
    
    \begin{scope}[xshift=0cm,yshift=-1.5cm,line width=2pt]
      \draw [dashed,Aluminium2] (-3.7,-3.5) rectangle ++(7.4,4);
      
      \node (lu0)     [anchor=center] at (0,0)    {\function{hlu}[\(A_{\clt_0,\clt_0},L_{\clt_0,\clt_0},U_{\clt_0,\clt_0}\)]};
      \node (htrsu)   [anchor=center]   at (-1.5,-0.75) {\function{htrsu}[\(U_{\clt_0,\clt_0},A_{\clt_1,\clt_0},L_{\clt_1,\clt_0}\)]};
      \node (htrsl)   [anchor=center]   at (1.5,-1.25)  {\function{htrsl}[\(L_{\clt_0,\clt_0},A_{\clt_0,\clt_1},U_{\clt_0,\clt_1}\)]};
      \node (Update)  [anchor=center] at (0,-2)   {\function{hmul}[\(-1,L_{\clt_1,\clt_0},U_{\clt_0,\clt_1},A_{\clt_1,\clt_1}\)]};
      \node (lu1)     [anchor=center] at (0,-3)   {\function{hlu}[\(A_{\clt_1,\clt_1},L_{\clt_1,\clt_1},U_{\clt_1,\clt_1}\)]};

      \draw [->] (lu0) -- (htrsu);
      \draw [->] (lu0) -- (htrsl);
      \draw [->] (htrsu) -- (Update);
      \draw [->] (htrsl) -- (Update);
      \draw [->] (Update) -- (lu1);
    \end{scope}
  \end{tikzpicture}
  \caption{Task refinement and resulting sub-tasks of \function{hlu}[\(A_{\clt,\clt},L_{\clt,\clt},U_{\clt,\clt}\)].}
  \label{fig:hlurefine}
\end{figure}

Due to the recursive nature of the \mcH-arithmetic functions, they will produce \emph{sub-tasks}, i.e., all subsequent
function calls within such an arithmetic function, which will \emph{replace} the original task. In
Figure~\ref{fig:hlurefine} this is shown for the function \function{hlu}.

For a task \(t\), let \(V_t\) be the set of sub-tasks. The tasks \(t' \in V_t\) will have a data dependency relation
between them, e.g., output data of one task is needed as the input of another task.

We can formalise these data dependencies in the context of \mcH-matrices with the help of the matrix blocks the
corresponding tasks work on. Each of these matrix blocks is identified by block index sets \(\clt \times \cls \in
T\). For the function \function{hlu} those blocks are \(A_{\clt,\clt}, L_{\clt,\clt}\) and \(U_{\clt,\clt}\), with input
data defined by \(A_{\clt,\clt}\) and output data defined by \(L_{\clt,\clt}\) and \(U_{\clt,\clt}\). For the
\mcH-arithmetic tasks we will identify these matrix blocks as a pair consisting of the corresponding block index set and
an identifier representing the (global) matrix, e.g., \(A, L\) or \(U\).

\begin{define}[Data Dependencies] \label{def:datadep}
  Let \(\mcI\) be a set of identifiers and let \function{id} denote the mapping of matrices to their identifiers. For
  each task \(t\) let \(t_{\idep} \subset \mcI \times V(T)\) denote the set of \emph{input} data dependencies and
  \(t_{\odep} \subset \mcI \times V(T)\) the set of \emph{output} data dependencies, respectively.
\end{define}

In Table~\ref{tbl:taskdep} the sets of input/output data dependencies is shown for the previously introduced \mcH-matrix
functions.

\begin{table*}
  \centering
  \begin{tabular}{lll}
    \multicolumn{1}{c}{Task}
    & \multicolumn{1}{c}{\(t_{\idep}\)}
    & \multicolumn{1}{c}{\(t_{\odep}\)} \\
    \toprule
    \function{hlu}[\(A_{\clt,\clt},L_{\clt,\clt},U_{\clt,\clt}\)]
    & \set{ (\function{id}[A], \clt \times \clt) }
    & \set{ (\function{id}[L], \clt \times \clt)), (\function{id}[U], \clt \times \clt) }
    \\
    \function{htrsl}[\(L_{\clt,\clt},M_{\clt,\cls},X_{\clt,\cls}\)]
    & \set{ (\function{id}[L], \clt \times \clt), (\function{id}[M], \clt \times \cls) }
    & \set{ (\function{id}[X], \clt \times \cls) }
    \\
    \function{htrsu}[\(U_{\clt,\clt},M_{s,t},X_{s,t}\)]
    & \set{ (\function{id}[U], \clt \times \clt), (\function{id}[M], \cls \times \clt) }
    & \set{ (\function{id}[X], \cls \times \clt) }
    \\
    \function{hmul}[\(A_{t,r},B_{r,s},C_{\clt,\cls}\)]
    & \set{ (\function{id}[A], \clt \times \clr), (\function{id}[B], \clr \times \cls) }
    & \set{ (\function{id}[C], \clt \times \cls) }
    \\
  \end{tabular}
  \caption{Input/output Dependencies for \mcH-LU tasks.}
  \label{tbl:taskdep}
\end{table*}

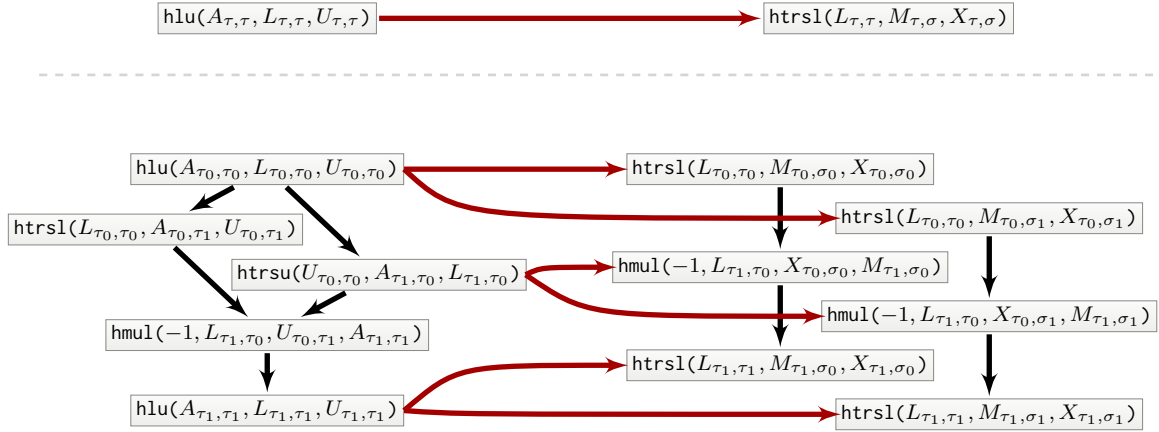
\begin{figure*}[htb]
  \centering
  \begin{tikzpicture}[line width=1pt,>=latex',every node/.style=task]

    \begin{scope}[line width=2pt]
      \node (lu0)    [anchor=center] at (0,0)   {\function{hlu}[\(A_{\clt,\clt},L_{\clt,\clt},U_{\clt,\clt}\)]};
      \node (htrsl)  [anchor=west]   at (6.5,0)  {\function{htrsl}[\(L_{\clt,\clt},M_{\clt,\cls},X_{\clt,\cls}\)]};
      \draw [ScarletRed3,->] (lu0) -- (htrsl);
    \end{scope}

    \draw [dashed,Aluminium2] (-3,-.75) -- (12,-.75);

    \begin{scope}[yshift=-2cm,line width=2pt]
      
      \node (lu0)     [anchor=center] at (0,0)   {\function{hlu}[\(A_{\clt_0,\clt_0},L_{\clt_0,\clt_0},U_{\clt_0,\clt_0}\)]};
      \node (htrsu)   [anchor=east]   at (.5,-0.8) {\function{htrsl}[\(L_{\clt_0,\clt_0},A_{\clt_0,\clt_1},U_{\clt_0,\clt_1}\)]};
      \node (htrsl)   [anchor=west]   at (-.5,-1.4)  {\function{htrsu}[\(U_{\clt_0,\clt_0},A_{\clt_1,\clt_0},L_{\clt_1,\clt_0}\)]};
      \node (Update)  [anchor=center] at (0,-2.2)  {\function{hmul}[\(-1,L_{\clt_1,\clt_0},U_{\clt_0,\clt_1},A_{\clt_1,\clt_1}\)]};
      \node (lu1)     [anchor=center] at (0,-3.2)  {\function{hlu}[\(A_{\clt_1,\clt_1},L_{\clt_1,\clt_1},U_{\clt_1,\clt_1}\)]};

      \draw [->] (lu0) -- (htrsu);
      \draw [->] (lu0) -- (htrsl);
      \draw [->] (htrsu) -- (Update);
      \draw [->] (htrsl) -- (Update);
      \draw [->] (Update) -- (lu1);
    \end{scope}

    
    \begin{scope}[xshift=8cm,yshift=-2cm,line width=2pt]

      \node (htrsl00)    [anchor=center] at (-1.25,0)   {\function{htrsl}[\(L_{\clt_0,\clt_0}, M_{\clt_0,\cls_0}, X_{\clt_0,\cls_0}\)]};
      \node (htrsl01)    [anchor=center] at (1.5,-0.65)    {\function{htrsl}[\(L_{\clt_0,\clt_0}, M_{\clt_0,\cls_1}, X_{\clt_0,\cls_1}\)]};
      \node (Update10)   [anchor=center] at (-1.25,-1.3)  {\function{hmul}[\(-1, L_{\clt_1,\clt_0}, X_{\clt_0,\cls_0}, M_{\clt_1,\cls_0}\)]};
      \node (Update11)   [anchor=center] at (1.5,-1.95)   {\function{hmul}[\(-1, L_{\clt_1,\clt_0}, X_{\clt_0,\cls_1}, M_{\clt_1,\cls_1}\)]};
      \node (htrsl10)    [anchor=center] at (-1.25,-2.6)  {\function{htrsl}[\(L_{\clt_1,\clt_1}, M_{\clt_1,\cls_0}, X_{\clt_1,\cls_0}\)]};
      \node (htrsl11)    [anchor=center] at (1.5,-3.25)   {\function{htrsl}[\(L_{\clt_1,\clt_1}, M_{\clt_1,\cls_1}, X_{\clt_1,\cls_1}\)]};
      
      \draw [->] (htrsl00) -- (Update10);
      \draw [->] (htrsl01) -- (Update11);
      \draw [->] (Update10) -- (htrsl10);
      \draw [->] (Update11) -- (htrsl11);
    \end{scope}

    \begin{scope}[yshift=-2cm,line width=2pt,color=ScarletRed3]
      \draw [->] (1.8,0) -- (4.75,0);
      \draw [->] (1.8,0) .. controls (2.5,-0.65) .. (7.55,-.65);
      \draw [->] (3.4,-1.4) .. controls (3.6,-1.3) .. (4.6,-1.3);
      \draw [->] (3.4,-1.4) .. controls (4.2,-1.95) .. (7.35,-1.95);
      \draw [->] (1.8,-3.2) .. controls (2.5,-2.6) .. (4.75,-2.6);
      \draw [->] (1.8,-3.2) .. controls (2.5,-3.25) .. (7.55,-3.25);
    \end{scope}
  \end{tikzpicture}
  \caption{Dependencies between parent tasks (top) and refined tasks (bottom).}
  \label{fig:deprefine}
\end{figure*}

Based on the data dependencies the task dependencies can be defined:

\begin{define}[Task Dependencies] \label{def:taskdep}
  Let \(t^i \ne t^j\) be two tasks. We say that \(t^i\) \emph{precedes} \(t^j\), written as \(t^i \tdep t^j\), iff
  \begin{displaymath}
    \exists (\operatorname{id}^i,b^i) \in t^i_{\odep},
    (\operatorname{id}^j,b^j) \in t^j_{\idep} :
    \operatorname{id}^i = \operatorname{id}^j \wedge b^i \cap b^j \ne \emptyset .
  \end{displaymath}
  Furthermore, for any task \(t\) let \(S_t \subseteq \mcT\) be the set of \emph{successors} of \(t\), e.g., \(S_t :=
  \set{ g : t\,\tdep g}\).
\end{define}

For the general case, we assume that the sub tasks \(V_t\) and the dependencies \(E_t \subset V_t \times V_t\) between
tasks in \(V_t\), forming a local graph \(G_t = ( V_t, E_t ) \), are user-provided for each task \(t\). Normally, these
directly follow from the definition of the standard \mcH-arithmetic functions, e.g., instead of a function call, a
sub-task is created (rf. Figure~\ref{fig:hlurefine}).

\begin{remark} \label{rem:loops}
  It is important that \(G_t\) must not include a loop in the corresponding task graph. Otherwise, the result of the
  task graph generation below will not produce a DAG, as is needed for the execution phase of the task graph.
\end{remark}

\begin{remark} \label{rem:autodeps}
  For many \mcH-matrix algorithms, including \mcH-LU, the construction of \(E_t\) can be automated by comparing the
  input/output data dependencies of the sub-tasks in \(V_t\), which further simplifies the whole process of task graph
  generation.
\end{remark}

After all tasks are refined and the sub-tasks together with their local dependencies are given, the next step is to set
up the task dependencies between sub-tasks of tasks \(t,g\) with \(t \tdep g\). This can be done automatically
using the data dependencies of the sub-tasks. Let \(V_t = \set{t_1,t_2}\). Then also \(t_1 \tdep g\) and \(t_2 \tdep g\)
holds. However, if \(g\) is refined, i.e., \(V_g = \set{g_1, g_2}\), the task dependencies \(t_1 \tdep g_1, t_1 \tdep
g_2, t_2 \tdep g_1\) and \(t_2 \tdep g_2\) do not necessarily apply. Therefore, when refining tasks and by that also
their dependencies, only those task dependencies as due to Definition~\ref{def:taskdep} will
remain. Algorithm~\ref{alg:subdeps} performs this comparison of sub-tasks to restrict the dependency set. An example of
the result for the \mcH-LU factorization is shown in Figure~\ref{fig:deprefine}.

\begin{algorithm}{Inheritance and refinement of sub-task dependencies}{alg:subdeps}
  \Procedure{refine\_sub\_deps}{in: $t$, out: $E$}
  \For{ \(g \in S_t \) }
    \If{ \(V_g \ne \emptyset\) } \(S := V_g\);
    \Else \hspace{5.95em} \(S := \set{g}\);
    \EndIf

    \For{ \(t' \in V_t, s \in S \) }
      \If{ \( t' \tdep s \) }
        \State \(E := E \cup \set{(t',s)}\);
      \EndIf
    \EndFor 
  \EndFor 
  \EndProcedure
\end{algorithm}

The same dependency refinement also has to be performed if the task \(t\) is not refined but \(g\) is, e.g., replacing
\(t \tdep g\) by \(\set{t \tdep g_1, t \tdep g_2}\). The corresponding algorithm works in an analog way to
Algorithm~\ref{alg:subdeps} and is shown in Algorithm~\ref{alg:locdeps}.

\begin{algorithm}{Refinement of local task dependencies}{alg:locdeps}
  \Procedure{refine\_loc\_deps}{in: $t$, out: $E$}
  \For{ \(g \in S_t \) }
    \If{ \(V_g \ne \emptyset\) } \(S := V_g\);
    \Else \hspace{5.95em} \(S := \set{g}\);
    \EndIf

    \For{ \(s \in S \) }
      \If{ \( t\,\tdep s \) }
        \State \(E := E \cup \set{(t,s)}\);
      \EndIf
    \EndFor 
  \EndFor 
  \EndProcedure
\end{algorithm}

\pgfdeclareimage[height=4cm]{dagA}{pics/dag_A}
\pgfdeclareimage[height=5cm]{dagLU}{pics/dag_LU}
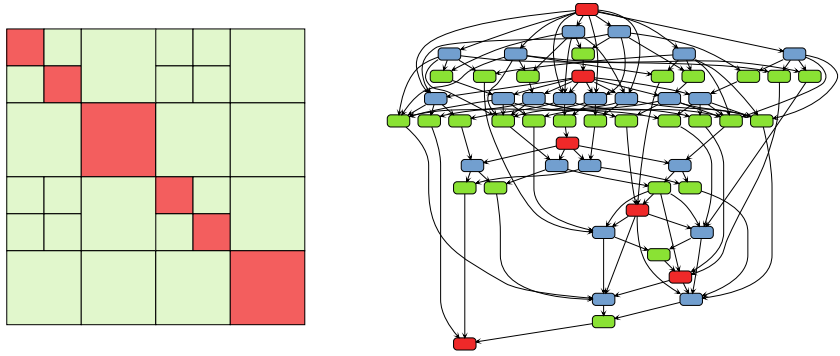
\begin{figure*}[htb]
  \centering
  \begin{tikzpicture}
    \draw (0,0) node [anchor=center] {\pgfuseimage{dagA}};
    \draw (6,0) node [anchor=center] {\pgfuseimage{dagLU}};
  \end{tikzpicture}
  \caption{\mcH-matrix (left) and corresponding DAG for \mcH-LU factorization (right).}
  \label{fig:dagLU}
\end{figure*}

For the computation of the task graph, both steps, e.g., task refinement and dependency refinement, are now put
together in an iterative process as is shown in Algorithm~\ref{alg:compdag}. In each step, first the current tasks are
refined (assuming user-provided sub-tasks and sub-task dependencies), followed by the refinement of the inherited
dependencies. If after both steps, a task was neither refined nor any of its successor task were, it will not change in
further iteration steps and may be removed from the workset of subsequent loops. If no task remains to be refined, the
iteration finishes. The number of iterations is given by \(\operatorname{depth}(\mcT)\). The start of the computation is
defined by the single task for the top-level call to the \mcH-arithmetic function, e.g., \function{hlu}[\(A,L,U\)].

\begin{remark} \label{rem:iterdepth}
  In practise, it may be more efficient to stop the iteration if the tasks are too small, e.g., if the overhead of
  handling the tasks outweighs the computation performed within the tasks. This may either be done by stopping the
  recursion before reaching \(\operatorname{depth}(\mcT)\) or by stopping the refinement of tasks at a user-specified
  matrix block size.
\end{remark}

\begin{algorithm}{Computation of task graph}{alg:compdag}
  \Procedure{compute\_dag}{in: $t$, out: $G = (V, E)$}
    \State \(N := \set{t}\); \(V := \emptyset\); \(E := \emptyset\);
    \While{ \(N \ne \emptyset \) }
      \ForAll{ \(g \in N\) }
        \State generate \(V_g, E_g\);
      \EndFor
      \State \(N' := \emptyset\);
      \ForAll{ \(g \in N\) }
        \If{ \(V_g = \emptyset\) }
          \State \(\tilde S_g := S_g\);
          \State \function{refine\_loc\_deps}[\(g\)];
          \If{ \(S_g \ne \tilde S_g \) }
            \State \(N' := N' \cup \set{g}\);
          \Else
            \State \(V := V \cup \set{g}\);
            \State \(E := E \cup \set{g} \times S_g\);
          \EndIf
        \Else
          \State \function{refine\_sub\_deps}[\(g\)];
          \State \(N' := N' \cup V_g\);
        \EndIf
      \EndFor
      \State \(N := N'\);
    \EndWhile
  \EndProcedure
\end{algorithm}

The result \(G = (V,E)\) of Algorithm~\ref{alg:compdag} is a DAG for the \mcH-arithmetic function. An example for the
\mcH-LU factorization is shown in Figure~\ref{fig:dagLU}. There, the red nodes correspond to the factorization of
diagonal matrix blocks. Off-diagonal matrix solves are colored blue while matrix updates are shown in green.

\subsection{Task graph with accumulators} \label{sec:accudag}

If accumulator-based arithmetic is used, the principles of task graph generation remain the same. Only the tasks and
their data dependencies will change, e.g., tasks for \function{add\_upd}, \function{shift\_upd} and
\function{apply\_upd} have to be generated according to Algorithms~\ref{alg:hluaccu} and \ref{alg:htrsmaccu}.

As for the data dependencies, the arithmetic functions for factorization and matrix solves depend now on the
accumulator of the matrix block (due to \function{shift\_upd} and \function{apply\_upd}). In contrast to the matrices
\(A, L\) and \(U\) these accumulators are distinct matrices, e.g., not being sub-blocks of each other. This leads to
identifiers in the data dependencies unique to each accumulator. Since \function{apply\_upd} modifies the actual matrix,
the identifier of the output data dependency is again the identifier of the global matrix. The dependency to the
accumulator of the parent matrix in \function{shift\_upd} and \function{apply\_upd} ensures the top-down hierarchy of
the application of updates via accumulators. Table~\ref{tbl:applydep} shows the (modified) data dependencies for the
corresponding tasks.

\begin{table*}
  \centering
  \begin{tabular}{lll}
    \multicolumn{1}{c}{Function}
    & \multicolumn{1}{c}{\(t_{\idep}\)}
    & \multicolumn{1}{c}{\(t_{\odep}\)}
    \\
    \toprule
    \function{hlu}[\(A_{\clt,\clt},L_{\clt,\clt},U_{\clt,\clt}\)]
    & \( \left\{ (\function{id}[A], \clt \times \clt), \right.\)
    & \set{ (\function{id}[L], \clt \times \clt)), (\function{id}[U], \clt \times \clt) }
    \\
    & \;\,\(\left. (\function{id}[\parent{A_{\clt,\clt}}], \clt \times \clt) \right\} \)
    &
    \\
    \function{htrsl}[\(L_{\clt,\clt},M_{\clt,\cls},X_{\clt,\cls}\)]
    & \( \left\{ (\function{id}[L], \clt \times \clt), (\function{id}[M], \clt \times \cls), \right. \)
    & \set{ (\function{id}[X], \clt \times \cls) }
    \\
    & \;\,\( \left. (\function{id}[\parent{M_{\clt,\cls}}],\clt \times \cls) \right\} \)
    &
    \\
    \function{htrsu}[\(U_{\clt,\clt},M_{\cls,\clt},X_{\cls,\clt}\)]
    & \( \left\{ (\function{id}[U], \clt \times \clt), (\function{id}[M], \cls \times \clt), \right. \)
    & \set{ (\function{id}[X], \cls \times \clt) }
    \\
    & \;\,\( \left. (\function{id}[\parent{M_{\cls,\clt}}], \cls \times \clt) \right\} \)
    &
    \\
    \midrule
    \function{add\_upd}[\(\alpha,A_{\clt,\clr},B_{\clr,\cls},C_{\clt,\cls}\)]
    & \set{ (\function{id}[A], \clt \times \clr), (\function{id}[B], \clr \times \cls) }
    & \set{ (\function{id}[C], \clt \times \cls), (\function{id}[\(C_{\clt,\cls}\)], \clt \times \cls) }
    \\
    \function{shift\_upd}[\(C_{\clt,\cls}\)]
    & \( \left\{ (\function{id}[\parent{C_{\clt,\cls}}], \clt \times \cls ), \right. \)
    & \set{ (\function{id}[\(C_{\clt,\cls}\)], \clt \times \cls ) }
    \\
    & \;\,\( \left. (\function{id}[\(C_{\clt,\cls}\)], \clt \times \cls ) \right\} \)
    &
    \\
    \function{apply\_upd}[\(C_{\clt,\cls}\)]
    & \( \left\{ (\function{id}[\parent{C_{\clt,\cls}}], \clt \times \cls ), \right. \)
    & \set{ (\function{id}[\(C\)], \clt \times \cls ) }
    \\
    & \;\,\( \left. (\function{id}[\(C_{\clt,\cls}\)], \clt \times \cls ) \right\} \)
    &
  \end{tabular}
  \caption{Input/Output Dependencies for \mcH-LU functions using accumulators.}
  \label{tbl:applydep}
\end{table*}

In Figure~\ref{fig:dagLUaccu} the task graph for the \mcH-LU factorization with accumulators is shown. The tasks for
applying updates are marked yellow, while factorization tasks and matrix solve tasks are again red and blue,
respectively. The green update tasks in Figure~\ref{fig:dagLU} are replaced by (equally colored) tasks for
\function{add\_upd}.

\pgfdeclareimage[height=5.5cm]{dagLUaccu}{pics/dag_LU_accu}
\begin{figure}
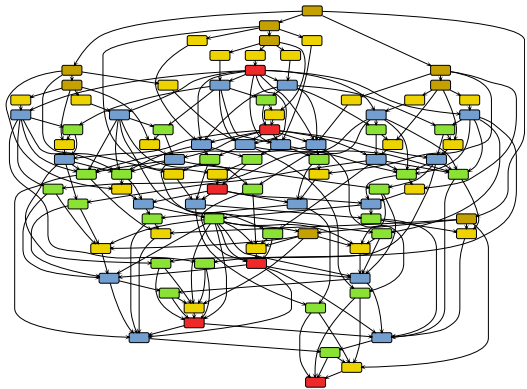

  \centering
  \pgfuseimage{dagLUaccu}
  \caption{\mcH-LU-DAG with accumulators.}
  \label{fig:dagLUaccu}
\end{figure}

\section{Optimization Techniques} \label{sec:opttech}

The above introduced task graph generation algorithm provides room for further optimization, where the goals are
improved memory requirements (Sections~\ref{sec:sparsify}) and runtime
(Section~\ref{sec:pardag}). Section~\ref{sec:mergeaccudag} shows an alternative way to incorporate accumulator
arithmetic into standard \mcH-arithmetic, thereby also reducing the computational cost of task graph generation.

\subsection{Edge Sparsification} \label{sec:sparsify}

During dependency refinement, the relation \(\tdep\) may result in unnecessary edges in \(G\), e.g. edges \((t,g)
\in E\) for tasks reachable by paths \(t=t_1,t_2,\ldots,t_{\ell}=g, (t_i,t_{i+1}) \in E, 1 \le i \le \ell-1\). Often
this is induced by the hierarchy of the \mcH-matrix.

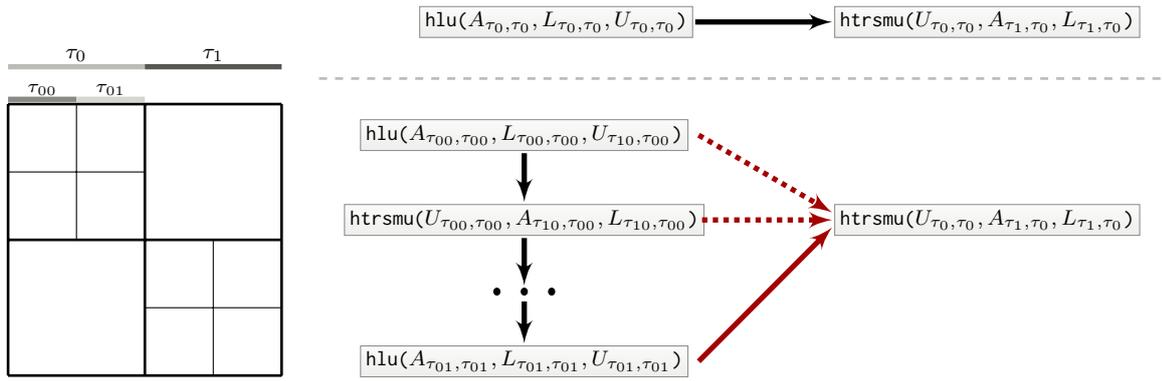
\begin{figure*}[htb]
  \centering
  \begin{tikzpicture}[xscale=0.9,yscale=-0.9]
    \draw (0,0) grid ++(2,2);
    \draw (2,2) grid ++(2,2);

    \begin{scope}[scale=2]
      \draw [line width=1pt] (0,0) grid ++(2,2);
    \end{scope}


    \draw (1,-0.5) node [anchor=south] {\footnotesize\(\clt_0\)}; \draw [Aluminium3,line width=2pt] (0,-0.55) -- ++(2,0);
    \draw (3,-0.5) node [anchor=south] {\footnotesize\(\clt_1\)}; \draw [Aluminium5,line width=2pt] (2,-0.55) -- ++(2,0);
    \draw (0.5,0) node [anchor=south] {\scriptsize\(\clt_{00}\)}; \draw [Aluminium4,line width=2pt] (0,-2pt) -- ++(1,0);
    \draw (1.5,0) node [anchor=south] {\scriptsize\(\clt_{01}\)}; \draw [Aluminium2,line width=2pt] (1,-2pt) -- ++(1,0);
  \end{tikzpicture}
  \quad
  \begin{tikzpicture}[xscale=0.9,yscale=1,line width=1pt,>=latex',every node/.style=task]
    \node (solve2)  [anchor=east] at (-1,1.5) {\function{hlu}[\(A_{\clt_0,\clt_0}, L_{\clt_0,\clt_0}, U_{\clt_0,\clt_0}\)]};
    \node (update3) [anchor=west] at (1,1.5) {\function{htrsmu}[\( U_{\clt_0,\clt_0}, A_{\clt_1,\clt_0}, L_{\clt_1,\clt_0}\)]};

    \draw [dashed,Aluminium3] (-6.5,0.75) -- (6,0.75);
    
    \node (lu13)    [anchor=center] at (-3.5,0) {\function{hlu}[\(A_{\clt_{00},\clt_{00}}, L_{\clt_{00},\clt_{00}}, U_{\clt_{10},\clt_{00}}\)]};
    \node (solve14) [anchor=center] at (-3.5,-1.125) {\function{htrsmu}[\(U_{\clt_{00},\clt_{00}}, A_{\clt_{10},\clt_{00}}, L_{\clt_{10},\clt_{00}}\)]};
    \node [draw=white,top color=white,bottom color=white] (dummy)   [anchor=center] at (-3.5,-2.1) {};
    \node (lu15)    [anchor=center] at (-3.5,-3.00) {\function{hlu}[\(A_{\clt_{01},\clt_{01}}, L_{\clt_{01},\clt_{01}}, U_{\clt_{01},\clt_{01}}\)]};

    \fill (-3.9,-2.075) circle (1.5pt);
    \fill (-3.5,-2.075) circle (1.5pt);
    \fill (-3.1,-2.075) circle (1.5pt);
    
    \node (update30) [anchor=west] at (1,-1.125) {\function{htrsmu}[\( U_{\clt_0,\clt_0}, A_{\clt_1,\clt_0}, L_{\clt_1,\clt_0}\)]};
    
    \begin{scope}[line width=2pt]
      \draw [->] (solve2) -- (update3);
      \draw [->] (lu13) -- (solve14);
      \draw [->] (solve14) -- (dummy);
      \draw [->] (dummy) -- (lu15);

      \draw [dotted,ScarletRed3,->] (-0.95,0)     -- (1.0,-1.025);
      \draw [dotted,ScarletRed3,->] (-0.9,-1.125) -- (1.0,-1.125);
      \draw [ScarletRed3,->] (-0.95,-3.0)  -- (1.0,-1.225);
      

      
    \end{scope}
  \end{tikzpicture}
  \caption{Generation of redundant edges (dotted red) during task refinement.}
  \label{fig:edges}
\end{figure*}

An example of this is shown in Figure~\ref{fig:edges}. There, the off-diagonal matrix solve of block \(A_{\clt_0,\clt_1}\)
depends on the factorization of block \(A_{\clt_0,\clt_0}\). During task refinement all sub-tasks of the factorization form a
dependency for the matrix solve task. However, since the factorization of \(A_{\clt_0,\clt_0}\) is only finished with the
factorization of \(A_{\clt_{01},\clt_{01}}\), only the dependency from this task is needed.

Another source of unnecessary edges might be the use of automatic task dependency generation for local sub-tasks (see
Remark~\ref{rem:autodeps}).

Though these redundant edges have no influence on the correctness of the DAG in terms of execution precedence, they
increase the number of edges of the DAG and by this its memory requirements. Furthermore, the runtime of the task graph
generation is higher since more edges have to be processed.

During task and dependency refinement, redundant edges are not generated between arbitray nodes in \(G\) since
refinement only affects neighbours of the corresponding tasks or of the sub-tasks. Therefore, a reachability test
between a task and nodes in its neighbourhood after a refinement step can detect such unneeded edges, e.g., if a path
\(t,\ldots,g\) of length at least two exists between \(t\) and \(g\), the edge \((t,g) \in E\) can be removed from the
graph. This is implemented in Algorithm~\ref{alg:sparsify} where for a node \(t\) (or all its sub-nodes) all non-direct
descendants, reachable within a given neighbourhood are determined. If for such a descendant \(s\) also an edge
\((t,s)\) exists, this edge will be removed.

\begin{algorithm}{Remove redundant edges}{alg:sparsify}
  \Procedure{remove\_redundant}{in: $t, N$, inout: $S_t$}
  \State \(A_t := \)\;\function{bfs\(_2\)}[\(t,G|_N\)];
  \ForAll{\(s \in S_t\)}
  \If{\(s \in A_t\)} \(S_t := S_t \setminus \set{s}\);
  \EndIf
  \EndFor
  \EndProcedure
  \Procedure{sparsify}{in: $t$}
  \If{\(V_t \ne \emptyset\)} \(N := V_t\);
  \Else \hspace{1.55cm} \(N := \set{t}\);
  \EndIf
  \ForAll{ \(s \in S_t\) }
  \If{ \(V_s \ne \emptyset \) } \(N := N \cup V_s\);
  \Else \hspace{1.88cm} \(N := N \cup \set{s}\);
  \EndIf
  \EndFor
  \If{\(V_t \ne \emptyset\)} 
  \ForAll{ \(t' \in S_t\) }
  \State \function{remove\_redundant}[\(t', N, S_{t'}\)];
  \EndFor
  \Else
  \State \function{remove\_redundant}[\(t, N, S_t\)];
  \EndIf
  \EndProcedure
\end{algorithm}

The neighbourhood is determined by the task and its successor tasks (or their sub-tasks). In case of a refined task, all
sub-tasks and their successors define the possible sub-graph to look for redundant edges.

\begin{remark}
  The function \function{bfs\(_2\)}[\(t,G'\)] returns all nodes visited by a breadth-first search starting at \(t\)
  within the graph \(G'\) with path lengths at least two.
\end{remark}

\begin{remark}
  In practise, the search for descendants in Algorithm~\ref{alg:sparsify} may further be limited by a maximal path
  length for efficiency reasons, thereby trading runtime with a slightly larger edge set. As an example, for standard
  \mcH-LU (without accumulators), already a path length of two resulted in a minimal edge set.
\end{remark}

In Figure~\ref{fig:LUsparse} an example of a task graph before and after removal of redundant edges is shown. There, the
number of edges is reduced from 74 to 54.

\pgfdeclareimage[height=5cm]{dagLUsparse}{pics/dag_LU_sparse}
\begin{figure}
  \centering
  \begin{tikzpicture}
    \draw (6.5,0) node [anchor=center] {\pgfuseimage{dagLUsparse}};
  \end{tikzpicture}
  \caption{\mcH-LU-DAG without redundant edges.}
  \label{fig:LUsparse}
\end{figure}
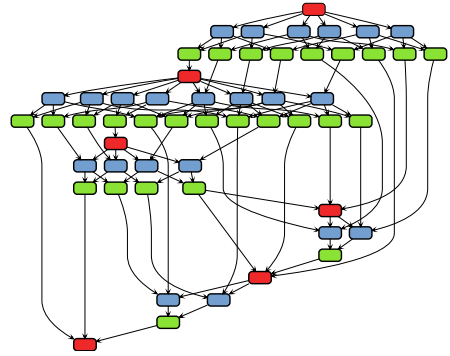

However, the removal of edges is a heuristical procedure since it may remove important edges, needed to guarantee data
dependencies in refined tasks.

\begin{figure*}[htb]
  \centering
  \begin{tikzpicture}[line width=1pt,>=latex',every node/.style=task]
    \begin{scope}[xshift=0cm,yshift=1.5cm,line width=2pt]
      \node (shift)   [anchor=center] at (0,1)  {\function{shift\_upd}[\(A_{\clt,\clt}\)]};
      \node (lu0)     [anchor=center] at (0,0)  {\function{hlu}[\(A_{\clt_0,\clt_0},L_{\clt_0,\clt_0},U_{\clt_0,\clt_0}\)]};
      \node (htrsu)   [anchor=center] at (0,-1) {\function{htrsu}[\(U_{\clt_0,\clt_0},A_{\clt_1,\clt_0},L_{\clt_1,\clt_0}\)]};

      \draw [->,ScarletRed3]        (shift) -- (lu0);
      \draw [->,ScarletRed3,dotted] (shift) .. controls (-2,1) and (-2.5,-0) .. (-1.9,-0.8);
      \draw [->]                    (lu0) -- (htrsu);
    \end{scope}

    \draw [dashed,Aluminium2] (2.75,3) -- ++(0,-4.3);
    
    \begin{scope}[xshift=7.5cm,yshift=1.5cm,line width=2pt]
      \node (shift0)   [anchor=center] at (0,1)  {\function{shift\_upd}[\(A_{\clt,\clt}\)]};
      \node (shift1)   [anchor=center] at (0,0.1)  {\function{shift\_upd}[\(A_{\clt_0,\clt_0}\)]};
      \node (lu00)     [anchor=center] at (0,-0.8)  {\function{hlu}[\(A_{\clt_{0_0},\clt_{0_0}},L_{\clt_{0_0},\clt_{0_0}},U_{\clt_{0_0},\clt_{0_0}}\)]};
      \node (shift2)   [anchor=center] at (0,-1.6)  {\function{shift\_upd}[\(A_{\clt_1,\clt_0}\)]};
      \node (htrsu00)  [anchor=center] at (0,-2.5) {\function{htrsu}[\(U_{\clt_{0_0},\clt_{0_0}},A_{\clt_{1_1},\clt_{0_0}},L_{\clt_{1_1},\clt_{0_0}}\)]};

      \draw [->,ScarletRed3] (shift0) -- (shift1);
      \draw [->,ScarletRed3] (shift1) -- (lu00);
      \draw [->,ScarletRed3,dotted] (shift0) .. controls (-3,1) and (-3,-1.6) .. (shift2);
      \draw [->,ScarletRed3] (shift2) -- (htrsu00);
      \draw [->] (1.5,-1.05) -- (1.5,-2.25);
    \end{scope}
  \end{tikzpicture}
  \caption{Edges induced by accumulator handling (red) in \mcH-LU before (left) and after (right) refinement. The right
    dotted edge will not be created if left dotted edge would have been removed by sparsification.}
  \label{fig:edgesprob}
\end{figure*}
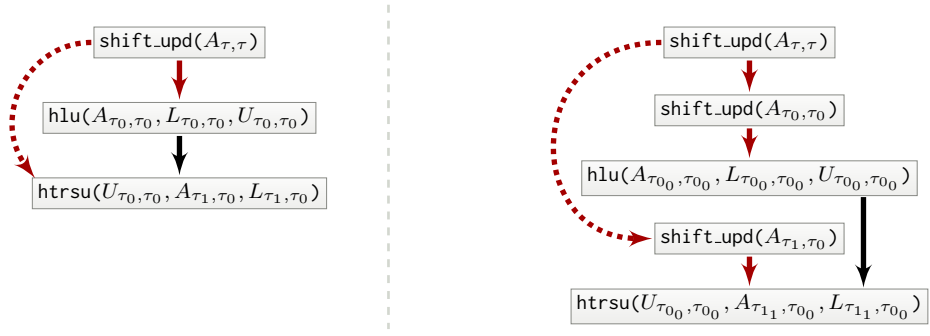

The task graph for the accumulator based arithmetic is such a negative example. There, the function \function{hlu} generates
sub-tasks for the shifting of updates applied to the current matrix block and factorization, matrix solves and updates
for its sub-blocks (see Figure~\ref{fig:edgesprob}). Since the tasks for \function{shift\_upd} have data dependencies
only in terms of the accumulator matrices, a corresponding \function{shift\_upd} predecessor node is needed to guarantee
that refined nodes will maintain the data dependencies needed for applying the accumulator updates.

\subsection{Parallel DAG Computation} \label{sec:pardag}

Algorithm~\ref{alg:compdag} has two major loops, first the refinement of the tasks and afterwards the refinement of the
dependencies. Both loops permit parallel execution as all performed operations are fully independent only affecting
local data. The result is shown in Algorithm~\ref{alg:parcompdag} (see also Algorithm~\ref{alg:compdag} for the omitted
parts).

\begin{algorithm}{Parallel computation of task graph}{alg:parcompdag}
  \Procedure{par\_compute\_dag}{in: $t$, out: $G = (V, E)$}
    \State \(N := \set{t}\); \(V := \emptyset\); \(E := \emptyset\);
    \While{ \(N \ne \emptyset \) }
      \ParForAll{ \(g \in N\) }
        \State generate \(V_g, E_g\);
      \EndFor
      \State \(N' := \emptyset\);
      \ParForAll{ \(g \in N\) }
        \If{ \(V_g = \emptyset\) }
          \State refine local dependencies;
        \Else
          \State refine dependencies of sub-tasks;
        \EndIf
      \EndFor
      \State \(N := N'\);
    \EndWhile
  \EndProcedure
\end{algorithm}

The situation changes if edge sparsification is applied. For a node \(t\) all non-local nodes, i.e., nodes not in
\(V_t\), in the neighbourhood used to find paths must remain unchanged during the optimization of the edge
set. Otherwise, the computation of the paths during the reachability test may result in undefined behaviour. To prevent
this, mutices associated with all tasks can be used, which are locked before and unlocked after
Algorithm~\ref{alg:sparsify} for all affected tasks as is shown in Algorithm~\ref{alg:sparsifymtx}.

\begin{algorithm}{Remove redundant edges with mutices}{alg:sparsifymtx}
  \Procedure{sparsify}{in: $t$}
  \State compute neighbourhood \(N\);
  \ForAll{ \(v \in N\) } lock(v);
  \EndFor
  \If{\(V_t \ne \emptyset\)} 
  \ForAll{ \(t' \in S_t\) }
  \State \function{remove\_redundant}[\(t', N, S_{t'}\)];
  \EndFor
  \Else
  \State \function{remove\_redundant}[\(t, N, S_t\)];
  \EndIf
  \ForAll{ \(v \in N\) } unlock(v);
  \EndFor
  \EndProcedure
\end{algorithm}



\begin{remark}
  In Algorithm~\ref{alg:parcompdag}, the details of the loop parallelization were left to the runtime system. In
  practise, it was more efficient to manually perform the splitting of the task set into separate chunks and perform the
  parallelization over the resulting set of chunks. If during task refinement such a chunk exceeds a predefined size, it
  is split into sub-chunks for the next iteration. Similarily, if chunks become too small due to the removal of finished
  nodes, they are joined with other (small) chunk sets.
\end{remark}

\begin{remark}
  The task graph generation from \cite{Kri:2013} is not so easily parallelizable as it has to follow the \mcH-matrix
  hierarchy to map the dependencies during \mcH-LU correctly. Furthermore, per matrix block only a very few tasks are
  generated, leaving also little room for parallelization.
\end{remark}

\subsection{Manually merging DAGs for accumulator arithmetic} \label{sec:mergeaccudag}

\begin{figure*}[htb]
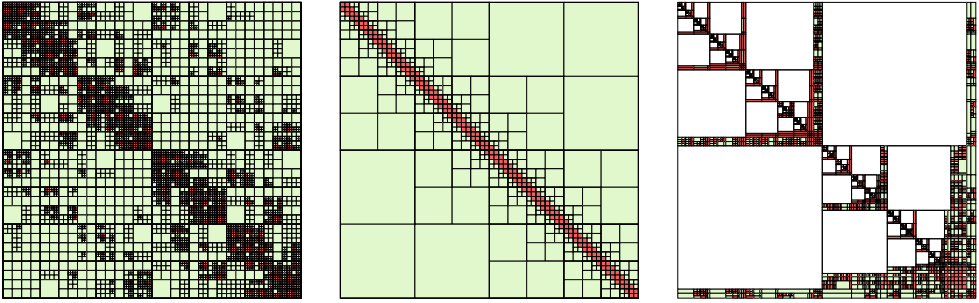

  \centering
  \pgfdeclareimage[width=4cm]{bctsphere}{pics/bct-sphere}
  \pgfdeclareimage[width=4cm]{bct1d}{pics/bct-1d}
  \pgfdeclareimage[width=4cm]{sparse}{pics/sparsealg_A}
  \pgfuseimage{bctsphere}
  \quad
  \pgfuseimage{bct1d}
  \quad
  \pgfuseimage{sparse}
  \caption{Block structure for model problems: Laplace SLP (left), 1D integral equation (middle) and sparse matrix (right).}
  \label{fig:bctmodel}
\end{figure*}

In Section~\ref{sec:accudag} the task graph was generated by following the \mcH-LU factorization and creating arithmetic
and accumulator tasks for the sub-blocks. The problem with this approach is that two different task graphs, one for the
accumulator handling and one for the standard \mcH-LU factorization, are created simultaneously. Because of this, more
nodes and edges have to be handled at the same time. Furthermore, edge sparsification is not possible (see
Section~\ref{sec:sparsify} and Figure~\ref{fig:edgesprob}).

An alternative approach is to first create only the task graph for \function{shift\_upd} and
\function{apply\_upd}. Afterwards the created accumulator tasks are used during the task graph construction for the
\mcH-LU factorization to explicitly create the dependencies between both graphs, e.g., add a dependency from a
\function{apply\_upd} task to a factorization task:
\begin{inlinealgorithm}
  \Procedure{hlu}{in:$A_{\clt,\clt},L_{\clt,\clt},U_{\clt,\clt}$}
  \If{\( (\clt,\clt) \not\in \mcL(T)\)}
  \State \(\ldots\)
  \Else
  \State\function{apply\_task}[\(A_{\clt,\clt}\)]\tdep\task{\(A_{\clt,\clt} = L_{\clt,\clt} U_{\clt,\clt}\)};
  \EndIf
  \EndProcedure
\end{inlinealgorithm}
Here, \function{apply\_task}[\(A_{\clt,\cls}\)] returns the pre-generated task for \function{apply\_upd} or \function{shift\_upd}
corresponding to the matrix block \(A_{\clt,\cls}\).

In an analog way, dependencies from \function{add\_upd} (which replaces the \function{hmul} call) to the corresponding
\function{apply\_upd}/\function{shift\_upd} task are created:
\begin{inlinealgorithm}
  \Procedure{hmul}{in:$\alpha,A_{t,r},B_{r,s},C_{\clt,\cls}$}
  \If{\(\set{(\clt,\clr),(\clr,\cls),(\clt,\cls)} \cap \mcL(T) = \emptyset \)}
  \State \(\ldots\)
  \Else
  \State \task{\function{add\_upd}[\(\alpha, A_{\clt,\clr}, B_{\clr,\cls}, C_{\clt,\cls}\)]} \tdep
  \State \hspace{.5cm} \function{apply\_task}[\(C_{\clt,\cls}\)];
  \EndIf
  \EndProcedure
\end{inlinealgorithm}

While this approach does not purely rely on the principle of data dependencies, it is faster since less edges are
processed during task graph generation.


\section{Numerical Experiments} \label{sec:numexp}

The new semi-automatic task-graph generation will be tested for several different \mcH-matrices, which differ by their
structure and dimension. For comparison, these test will also be performed for the DAG algorithm from
\cite{Kri:2013}, in the following referred to as the \emph{level-wise} method.

Please note, that only the generation of the task-graphs will be tested as the actual DAG execution does not differ
between the level-wise and the semi-automatic method. The reason for this is, that the tasks of the DAG are identical
and therefore also the computational work. In theory, a difference may exist due to overhead of the runtime system
scheduling the different task graphs. However, such a difference was not observed during the
experiments.

\begin{table}
  \centering
  \begin{tabular}{ll}
    \multicolumn{1}{c}{Software} 
    & \multicolumn{1}{c}{Version}
    \\
    \toprule
    \texttt{HLR} & 719c48f812e4 \\
    \texttt{HLIBpro} & 2.7.2 \\
    \texttt{GCC} & 8.2 \\
    \texttt{Intel TBB} & 2019.0 \\
    \texttt{Intel MKL} & 2018.4 \\
    \texttt{jemalloc} & 5.2.1 \\
  \end{tabular}
  \caption{Versions of software used for the experiments}
  \label{tab:software}
\end{table}

The versions of the different software used in the tests is shown in Table~\ref{tab:software}. All tests were performed
on a system with two Intel Xeon Gold 6148 CPUs and 192GB of main memory running SLES12 SP4.

\begin{remark}
  All tests were executed ten times for the same problem. Results in tables will show the \emph{median} of these
  results. The diagrams will also use the median for the corresponding plot and will furthermore show the worst/best
  result as a colored area.
\end{remark}

\subsection{Model Problems}

\pgfdeclareimage[width=7.5cm]{seqtime}{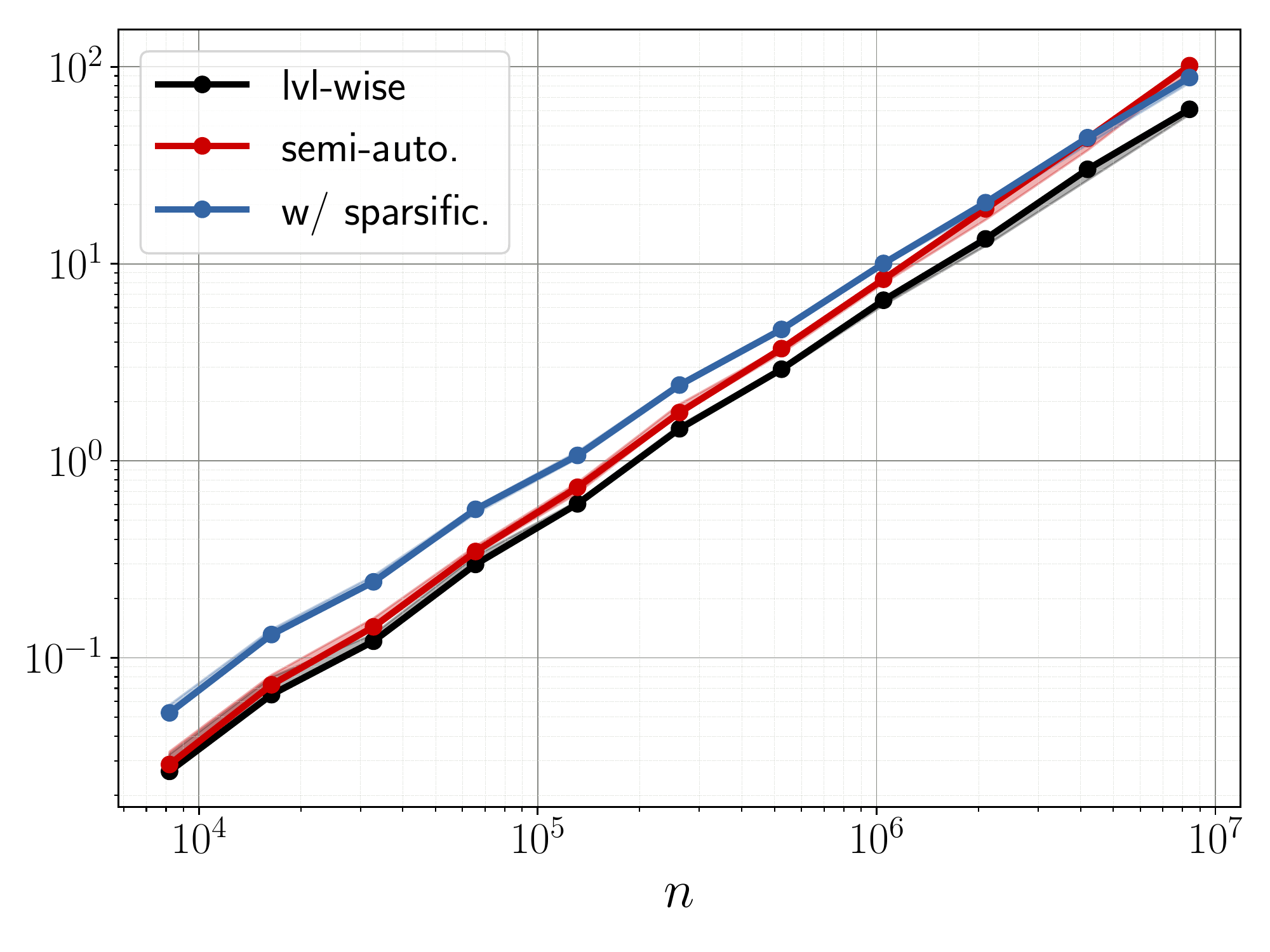}
\begin{figure*}[htb]
  \centering
  \begin{minipage}{0.45\linewidth}
    \pgfuseimage{seqtime}
  \end{minipage}
  \hspace{.1cm}
  {\begin{minipage}{0.45\linewidth}\small
    \begin{tabular}{rlll}
      \multicolumn{1}{c}{\(\mathbf{n}\)}
      & \multicolumn{1}{c}{\textbf{level-wise}}
      & \multicolumn{1}{c}{\textbf{semi-auto.}}
      & \multicolumn{1}{c}{\textbf{w/ sparsific.}} \\
      \toprule
      8.192     & \(2.47 \cdot 10^{-2}\) s & \(2.91 \cdot 10^{-2}\) s & \(5.43 \cdot 10^{-2}\) s \\
      16.384    & \(6.71 \cdot 10^{-2}\) s & \(7.34 \cdot 10^{-2}\) s & \(1.35 \cdot 10^{-1}\) s \\
      32.768    & \(1.22 \cdot 10^{-1}\) s & \(1.45 \cdot 10^{-1}\) s & \(2.49 \cdot 10^{-1}\) s \\
      65.536    & \(3.00 \cdot 10^{-1}\) s & \(3.53 \cdot 10^{-1}\) s & \(5.91 \cdot 10^{-1}\) s \\
      131.072   & \(6.12 \cdot 10^{-1}\) s & \(7.19 \cdot 10^{-1}\) s & \(1.10 \cdot 10^{0}\) s \\
      262.144   & \(1.48 \cdot 10^{0}\) s  & \(1.79 \cdot 10^{0}\) s  & \(2.50 \cdot 10^{0}\) s \\
      524.288   & \(2.94 \cdot 10^{0}\) s  & \(3.68 \cdot 10^{0}\) s  & \(4.80 \cdot 10^{0}\) s \\
      1.048.576 & \(6.59 \cdot 10^{0}\) s  & \(8.43 \cdot 10^{0}\) s  & \(1.03 \cdot 10^{1}\) s \\
      2.097.152 & \(1.30 \cdot 10^{1}\) s  & \(1.90 \cdot 10^{1}\) s  & \(2.09 \cdot 10^{1}\) s \\
      4.194.304 & \(3.02 \cdot 10^{1}\) s  & \(4.36 \cdot 10^{1}\) s  & \(4.46 \cdot 10^{1}\) s \\
      8.388.608 & \(6.15 \cdot 10^{1}\) s  & \(1.02 \cdot 10^{2}\) s  & \(9.03 \cdot 10^{1}\) s
    \end{tabular}
    \vspace{.5cm}
  \end{minipage}}
  \caption{Sequential runtime of level-wise and semi-automatic task graph generation with and without edge sparsification.}
  \label{fig:seqruntime}
\end{figure*}

\pgfdeclareimage[width=7.5cm]{graphsize}{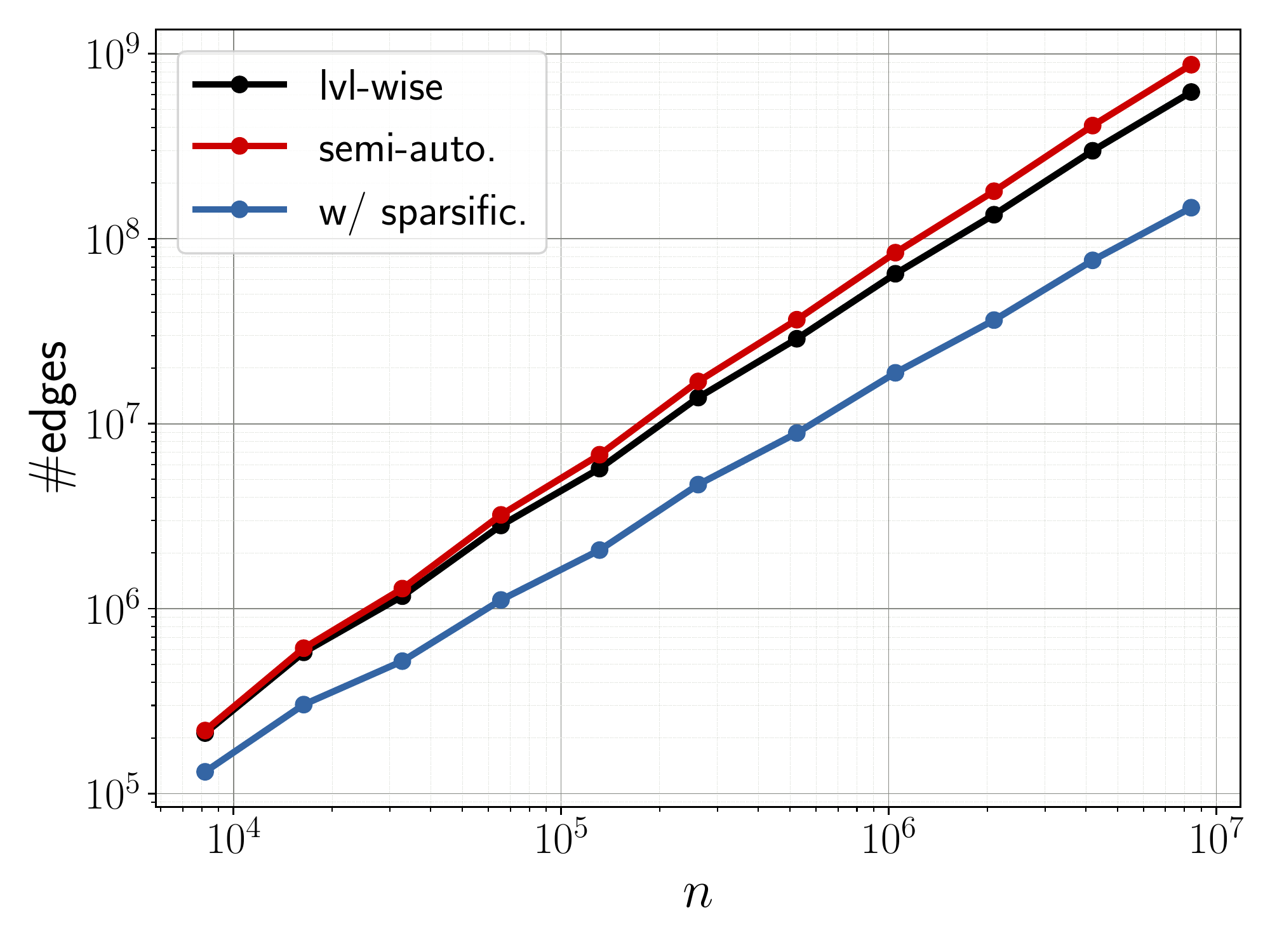}
\begin{table*}[htb]
  \centering
  \begin{tabular}{rrrrr}
    \multicolumn{1}{c}{\(\mathbf{n}\)}
    & \multicolumn{1}{c}{\textbf{\#nodes}}
    & \multicolumn{3}{c}{\textbf{\#edges}} \\
    &
    & \multicolumn{1}{c}{\textbf{level-wise}}
    & \multicolumn{1}{c}{\textbf{semi-auto.}}
    & \multicolumn{1}{c}{\textbf{w/ sparsific.}} \\
    \toprule
    8.192     &     38.653 &     212.698 &     219.494 &     131.288 \\
    16.384    &     87.346 &     580.992 &     611.866 &     303.133 \\
    32.768    &    150.139 &   1.169.104 &   1.284.578 &     520.694 \\
    65.536    &    321.362 &   2.819.304 &   3.214.902 &   1.115.069 \\
    131.072   &    597.784 &   5.721.682 &   6.798.604 &   2.074.099 \\
    262.144   &  1.346.326 &  13.848.468 &  16.926.586 &   4.687.193 \\
    524.288   &  2.556.413 &  28.798.228 &  36.502.620 &   8.904.326 \\
    1.048.576 &  5.370.314 &  64.666.356 &  84.095.200 &  18.844.125 \\
    2.097.152 & 10.351.022 & 134.903.680 & 180.723.520 &  36.323.779 \\
    4.194.304 & 21.699.437 & 298.927.488 & 409.007.162 &  76.419.056 \\
    8.388.608 & 41.824.170 & 621.029.914 & 872.616.702 & 147.203.171
  \end{tabular}
  \caption{Number of nodes and edges of the DAGs due to level-wise and semi-automatic task graph generation.}
  \label{tab:graphsize}
\end{table*}

The standard problem for the numerical examples is based on a boundary element discretization for the Laplace single
layer potential (Laplace SLP) while the domain is defined by the unit sphere:
\begin{equation} \label{eqn:slp}
  \int_{\Gamma} \frac{1}{\|x-y\|} u(x) dy = f(x), \quad x \in \Gamma
\end{equation}
with \(\Gamma = \set{ x \in \R^3 : \|x\|_2 = 1}\). Piecewise constant ansatz functions are used for the
discretization. Furthermore, standard admissibility
\begin{displaymath}
\min\left\{\op{diam}(t),\op{diam}(s)\right\} \le \eta \op{dist}(\clt,\cls)
\end{displaymath}
is applied for setting up the block tree.

\begin{remark}
  For all numerical examples, the matrix entries are not of importance as for the computation of the task graph, only
  the block tree is needed.
\end{remark}

The Laplace SLP model problem will be the default model problem for the numerical experiments below. If not stated
otherwise, the data from all figures and tables correspond to this problem.

While the block structure of the Laplace SLP problem resembles a typical \mcH-matrix block structure and therefore
serves as a reasonable approximate for other geometries, we will also consider the standard 1D model problem from
\cite{Boerm:2003b}:
\begin{equation} \label{eqn:1d}
  \int_0^1 \op{log} \|x-y\| u(x) dy = f(x), \quad x \in [0,1]
\end{equation}
Again, standard admissibility is used for the block tree, which results in a very coarse block structure of the
\mcH-matrix, corresponding to a very limited number of tasks per level. Therefore, the overhead due to refinement
is higher compared to the Laplace SLP example.

The two previous problems use boundary element methods to descretize an integral equation. The last model problem will instead
use the finite element method for the partial differential equation
\begin{equation} \label{eqn:pde}
  - \kappa \Delta u + b \cdot \nabla u = f \quad \text{in} \quad \Omega = ]0,1[^3.
\end{equation}
with a circular convection direction \(b(v_1,v_2,v_3):=(\frac{1}{2}-v_2,v_1-\frac{1}{2},0)^T\) and \(\kappa = 10^{-2}\). For the \mcH-matrix representation, algebraic
nested dissection clustering (see \cite{GrKrLeB08}) is used. The resulting block structure is different from the block
structure of the Laplace SLP and the 1D problem with a combination of large diagonal blocks, zero off-diagonal blocks
and rectangular blocks (see Figure~\ref{fig:bctmodel}).

\subsection{Comparing semi-automatic and level-wise DAG generation}

In Figure~\ref{fig:seqruntime} the sequential runtime of the level-wise and the semi-automatic algorithms are shown
together with the corresponding values for the sphere example.

As expected, the level-wise algorithm shows a faster runtime. The main reason for this is that the semi-automatic
approach is more compute intensive due to the many comparisons of data dependencies. Furthermore, the semi-automatic
algorithm has a significant management overhead due to memory allocation/deallocation of nodes and edges during task
refinement.

\begin{remark}
  This memory management overhead is also the reason why the memory allocation library \emph{jemalloc} \cite{jemalloc}
  was used as it resulted in a significant runtime improvement.
\end{remark}

Another reason for the slightly higher runtime is a larger number of edges as can be seen in
Table~\ref{tab:graphsize}. Though the number of nodes differs slightly since the level-wise approach uses additional
synchronization nodes for the diagonal factorization tasks, this difference is negligible (about 1--2\textperthousand). 


The number of edges can be decreased significantly by using edge sparsification from Section~\ref{sec:sparsify}. The
improvement of the results shown in Table~\ref{tab:graphsize} reach a factor of almost 6 at the largest problem
size. The number of edges with sparsification is also much smaller than with level-wise DAG construction.

Since edge sparsification involves additional computations, the runtime is normally increased. However, since also the
edge set is reduced, the computational savings due to this reduction finally lead to a faster runtime as can be seen in
Figure~\ref{fig:seqruntime}. 


\subsection{Parallel DAG generation}

\pgfdeclareimage[width=7.5cm]{timepar}{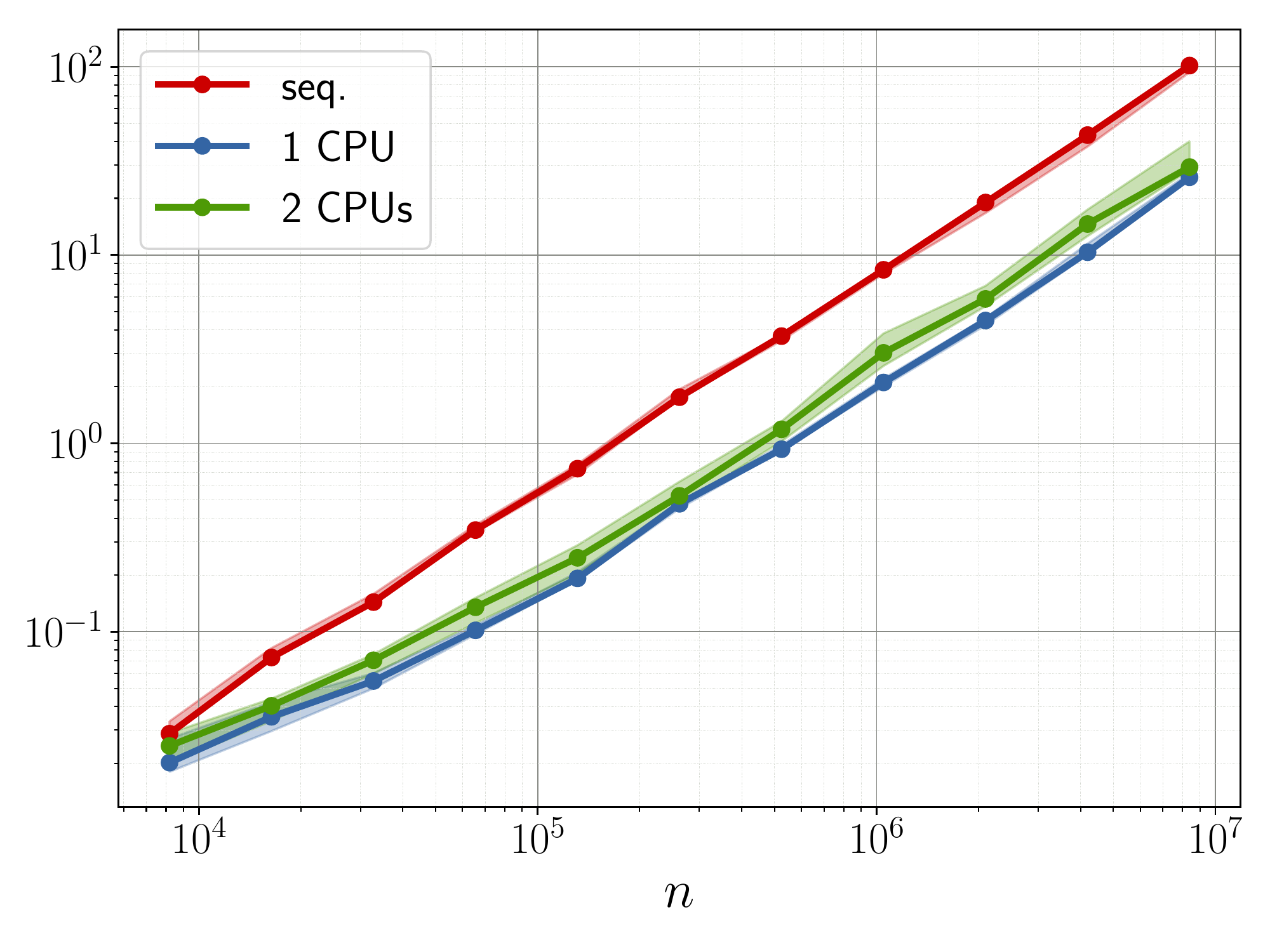}
\begin{figure*}[htb]
  \centering
  \begin{minipage}{0.45\linewidth}
    \pgfuseimage{timepar}
  \end{minipage}
  \hspace{1cm}
  {\begin{minipage}{0.35\linewidth}\small
    \begin{tabular}{rrr}
      \multicolumn{1}{c}{\(\mathbf{n}\)}
      & \multicolumn{2}{c}{\textbf{Parallel Speedup}}
      \\
      & \multicolumn{1}{c}{\textbf{1 CPU}}
      & \multicolumn{1}{c}{\textbf{2 CPUs}}
      \\
      \toprule
      8.192     & \speedup{2.908e-02}{2.053e-02} & \speedup{2.908e-02}{1.988e-02} \\
      16.384    & \speedup{7.336e-02}{3.390e-02} & \speedup{7.336e-02}{4.421e-02} \\
      32.768    & \speedup{1.451e-01}{4.906e-02} & \speedup{1.451e-01}{5.600e-02} \\
      65.536    & \speedup{3.526e-01}{1.070e-01} & \speedup{3.526e-01}{1.283e-01} \\
      131.072   & \speedup{7.186e-01}{1.947e-01} & \speedup{7.186e-01}{2.607e-01} \\
      262.144   & \speedup{1.789e+00}{4.556e-01} & \speedup{1.789e+00}{6.442e-01} \\
      524.288   & \speedup{3.678e+00}{9.320e-01} & \speedup{3.678e+00}{1.175e+00} \\
      1.048.576 & \speedup{8.431e+00}{2.168e+00} & \speedup{8.431e+00}{2.705e+00} \\
      2.097.152 & \speedup{1.903e+01}{4.518e+00} & \speedup{1.903e+01}{6.359e+00} \\
      4.194.304 & \speedup{4.328e+01}{1.041e+01} & \speedup{4.328e+01}{1.379e+01} \\
      8.388.608 & \speedup{1.016e+02}{2.623e+01} & \speedup{1.016e+02}{2.963e+01} \\
    \end{tabular}
    \vspace{.5cm}
  \end{minipage}}
  \caption{Parallel runtime using one CPU (20 cores) and two CPUs (40 cores).}
  \label{fig:timepar}
\end{figure*}

\pgfdeclareimage[width=7.5cm]{timeparsparse}{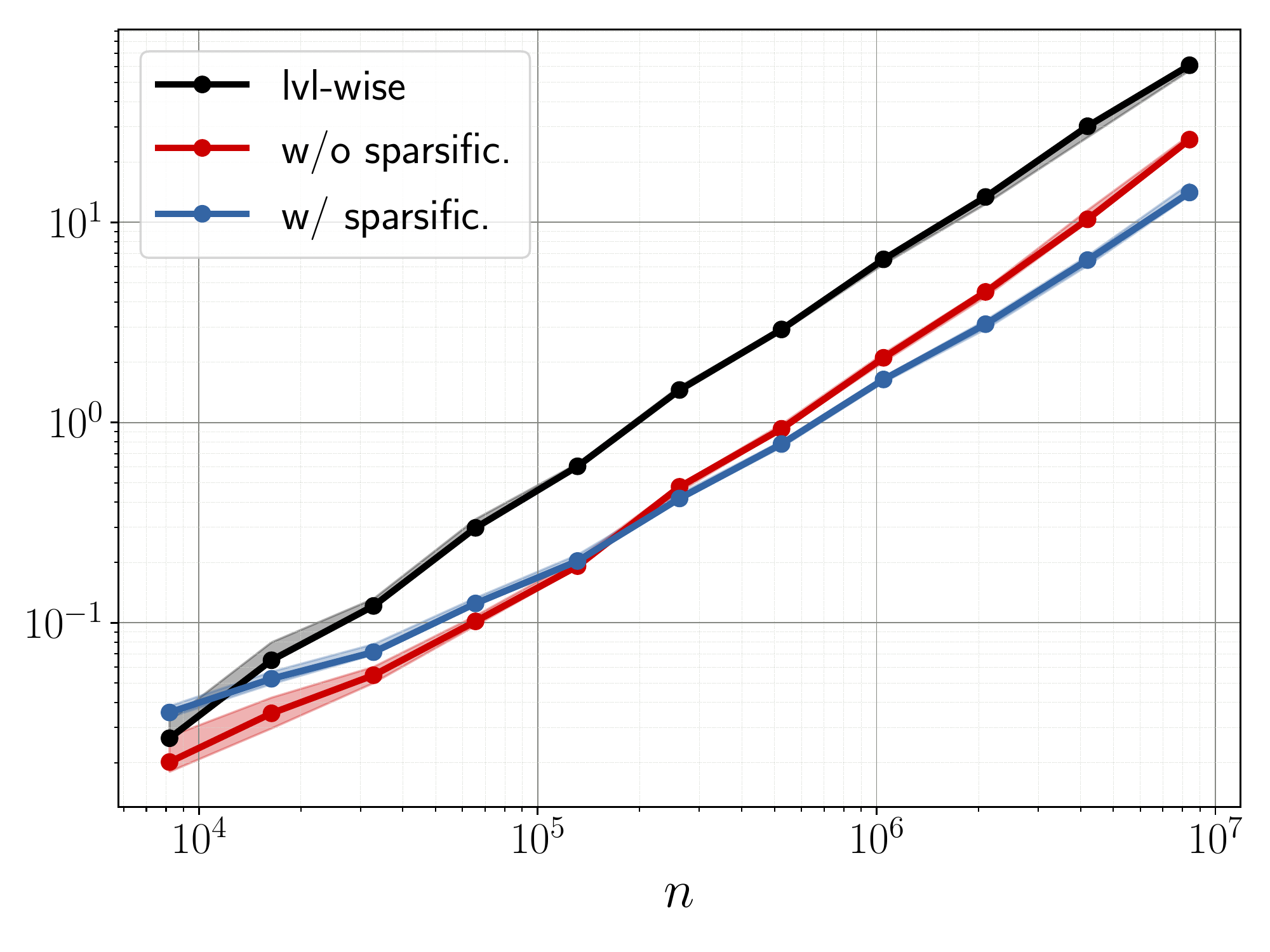}
\pgfdeclareimage[width=7.5cm]{speedup}{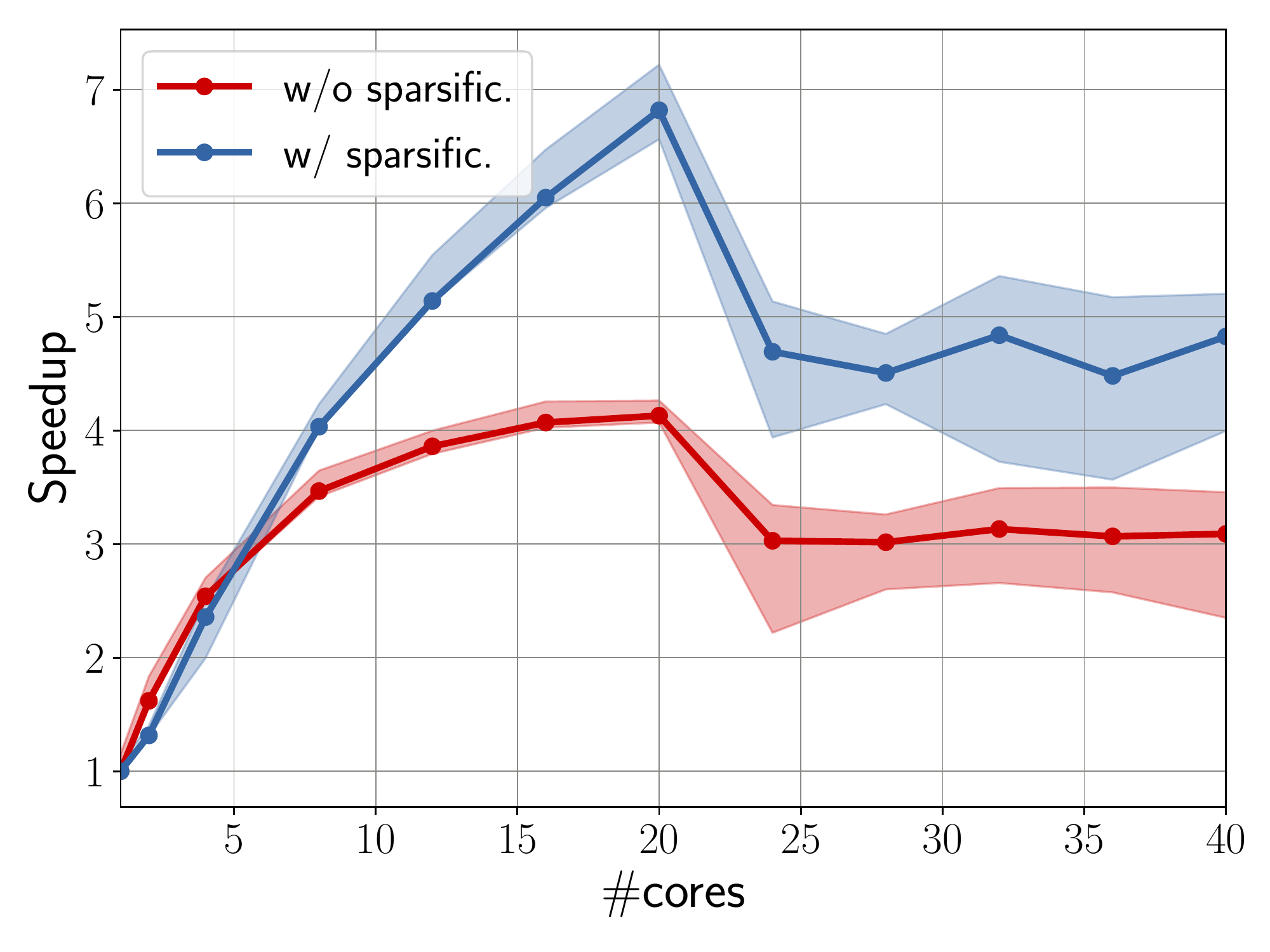}
\begin{figure*}[htb]
  \centering
  \pgfuseimage{timeparsparse}
  \quad
  \pgfuseimage{speedup}
  \caption{Parallel runtime (left) using one CPU (20 cores) and parallel speedup (right) for \(n=4.194.304\) with and
    without edge sparsification.}
  \label{fig:timeparspd}
\end{figure*}


The critical issue for the task graph generation is the low computational density of the computation coupled with mainly
indirect memory addressing using pointers as the graph data structure needs to be as flexible as possible. Furthermore,
the \mcH-matrices involved in the arithmetic need to be accessed simultaneously while generating the DAG, thereby
competing for memory bandwith. Therefore, the parallel scaling behaviour is not expected to be ideal.

The results shown in Figure~\ref{fig:timepar} confirm these expectations. The parallel speedup compared to the
sequential runtime is limited, achieving only a factor of 4 for a single CPU with 20 cores. When using two CPUs this
drops to a speedup of 3 due to more overhead, e.g., slower memory access for non-local data.

Nevertheless, the algorithm benefits from a parallel CPU and achieves maximal speedup already with a few number of CPU
cores as is shown in Figure~\ref{fig:timeparspd}, making the semi-automatic DAG generation faster on most computer systems
compared to the level-wise DAG generation.


When enabling edge sparsification, the same effect as in the sequential case can be observed, namely that for small
problem sizes the additional overhead leads to an increase in the runtime while the reduced number of edges finally
result in a faster algorithm. When comparing DAG generation with edge sparsification for sequential and parallel execution, the
parallel speedup is also higher as can be seen in Figure~\ref{fig:timeparspd}. This higher speedup is achieved
although additional mutices had to be used as explained in Section~\ref{sec:pardag}. Apparently the increase in
computational complexity per task due to the path search leads to a better usage of parallel resources.

\pgfdeclareimage[width=7.5cm]{onetime}{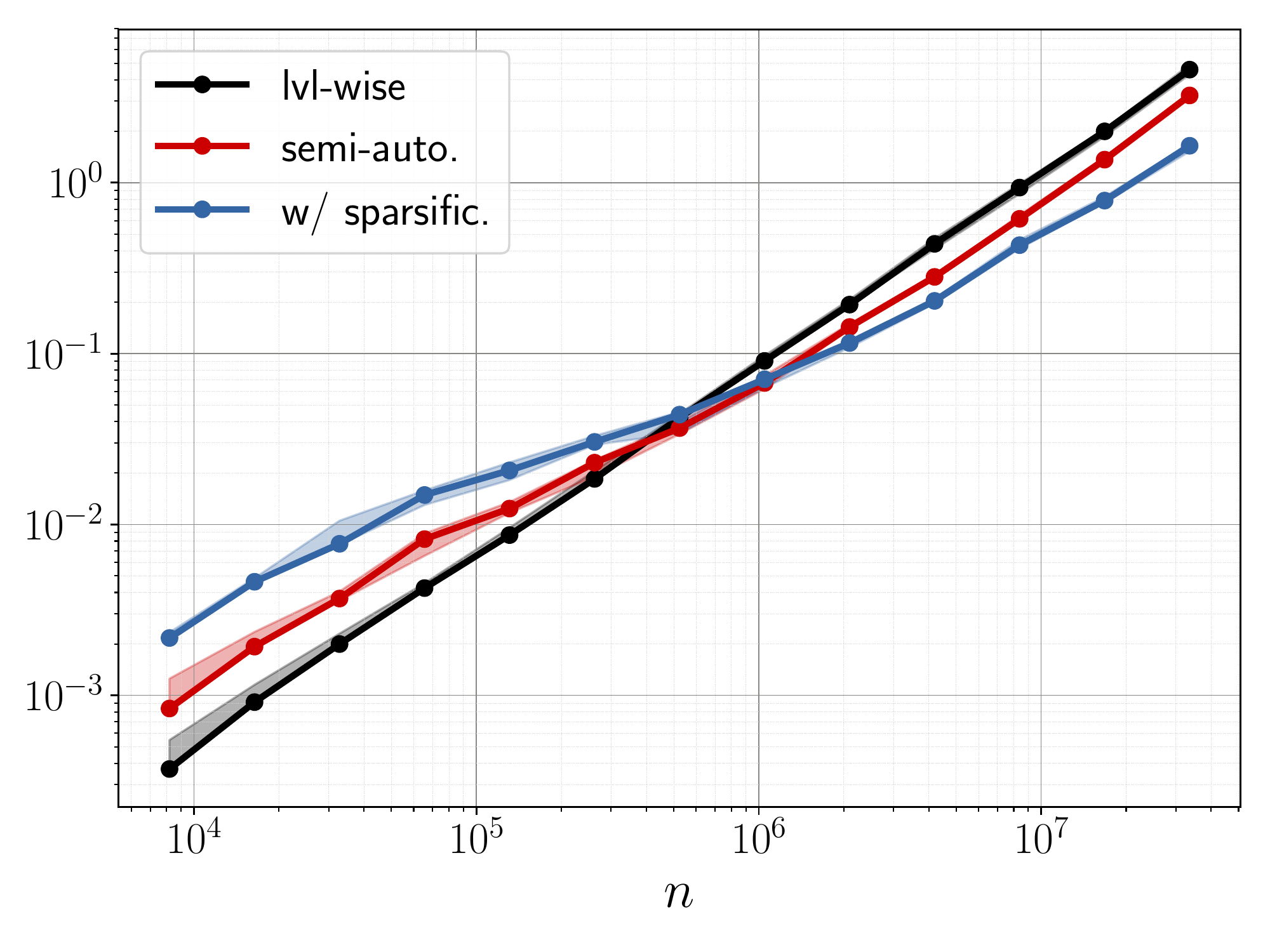}
\pgfdeclareimage[width=7.5cm]{sparsetime}{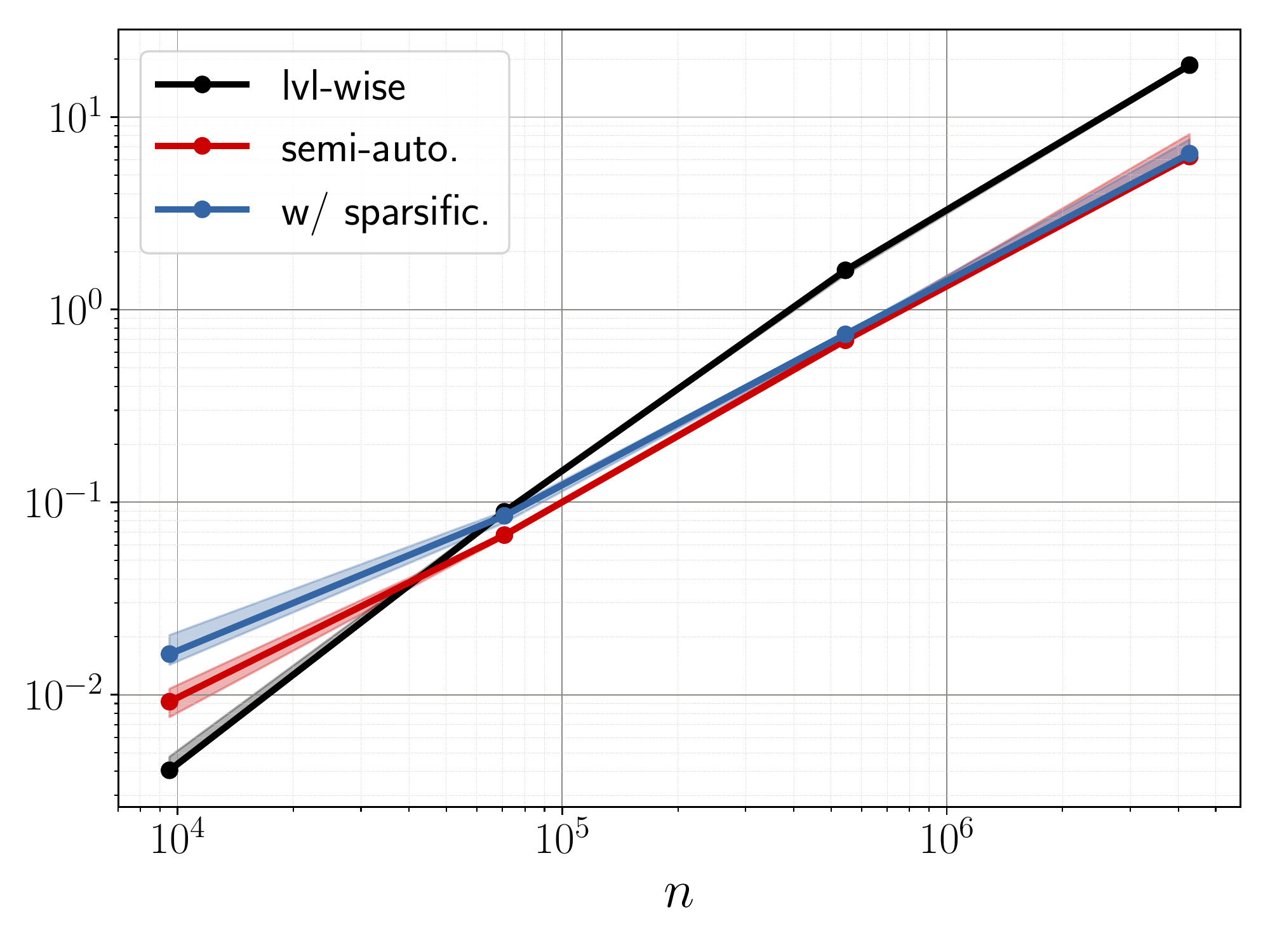}
\begin{figure*}[htb]
  \centering
  \pgfuseimage{onetime}
  \quad
  \pgfuseimage{sparsetime}
  \caption{Runtime for one CPU (20 cores) for 1D model problem \eqref{eqn:1d} (left) and PDE problem \eqref{eqn:pde} (right).}
  \label{fig:1d}
\end{figure*}

However, in both cases, a significant (sequential) overhead limits the achievable speedup, which is further
limited by using a second CPU due to a higher communication overhead. However, comparing the parallel runtime even with
a few CPU cores with the level-wise approach clearly shows an advantage of the semi-automatic method on practically all
computer systems nowadays. This is also shown in Figure~\ref{fig:percdag}. There, the runtime percentage of the task
graph generation on the full \mcH-LU factorization is shown for the Laplace SLP model problem on two CPUs (40 cores)
using the best runtime setup for creating the DAG. Since DAG execution scales much better compared to DAG construction,
the percentage is rather large. However, the runtime complexity of \mcH-LU is higher, leading to a smaller percentage
with larger problem sizes, even for the level-wise method. Furthermore, the semi-automatic approach is not only faster
compared to the old algorithm, but the relative portion does also shrink faster. Enabling edge sparsification further
reduces this part, albeit only for large problem sizes.

\begin{remark}
  For the DAG execution phase of the \mcH-LU factorization, the Laplace SLP example only needs a relatively small amount
  of floating point operations per index. For other problems, e.g. Helmholtz of Maxwell, the computational costs are
  much higher, further reducing the percentage of task graph generation on the full \mcH-LU factorization.
\end{remark}

\pgfdeclareimage[width=7.5cm]{percdag}{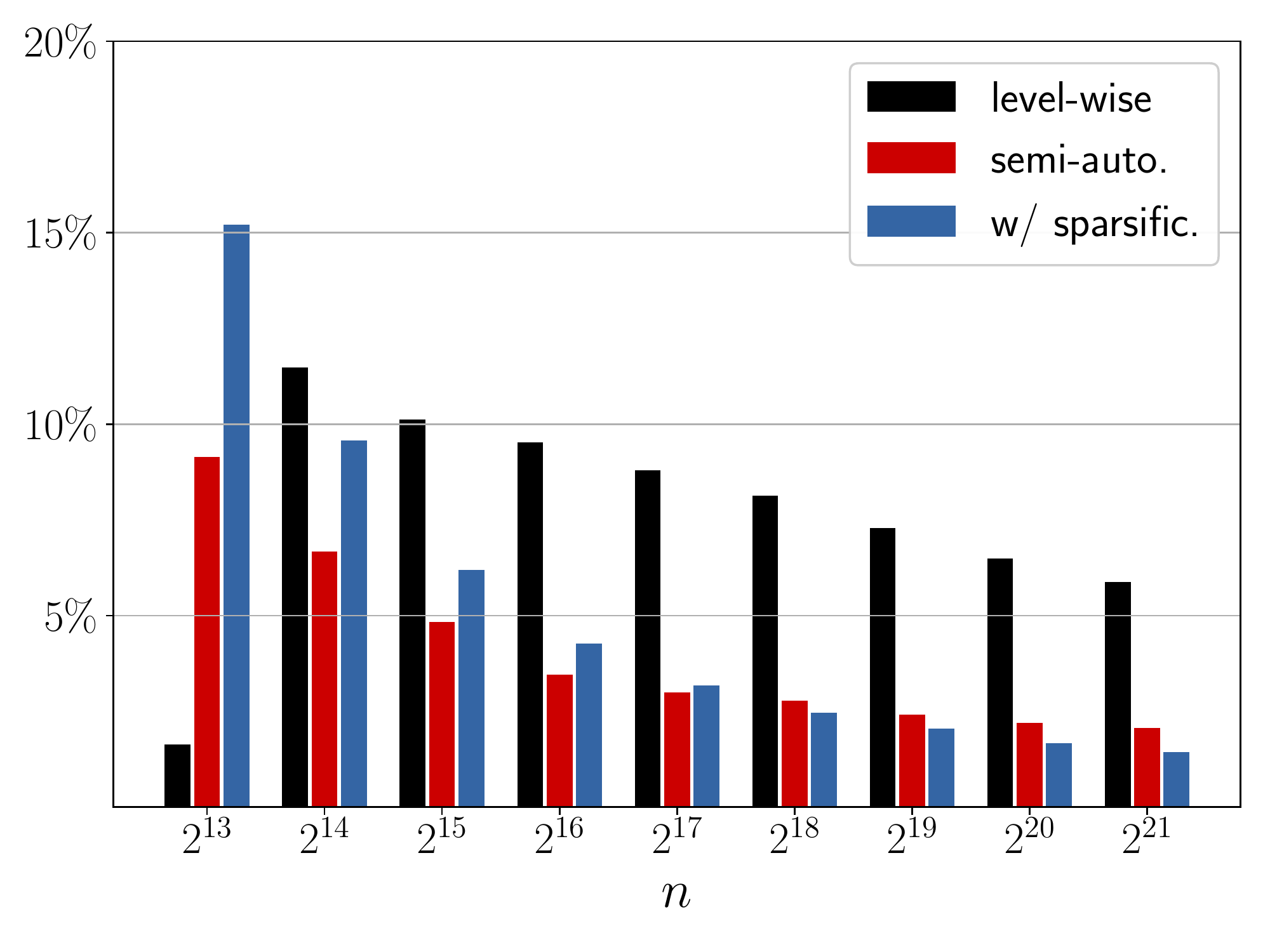}
\begin{figure}[htb]
  \centering
  \pgfuseimage{percdag}
  \caption{Percentage of task graph generation on full \mcH-LU algorithm}
  \label{fig:percdag}
\end{figure}


For the 1D model problem \eqref{eqn:1d} the general behaviour of the runtime and the number of edges is similar to the
Laplace SLP problem. However, due to the limited number of blocks per level and the deeper hierarchy of the \mcH-matrix,
the overhead of the semi-automatic task refinement is more pronounced. Therefore, the break-even point is achieved for
larger problem sizes and the advantage of the semi-automatic method (with or without edge sparsification) is smaller
compared to the level-wise methods.

The behaviour changes a little with the sparse matrix example \eqref{eqn:pde} as the percentage of the overhead of
the semi-automatic method is similar to the Laplace SLP problem. Furthermore, edge sparsification does not result in
a similar improvement as due to the sparse block structure, fewer edges per task are created in the first
place\footnote{For the largest problem size, the level-wise method resulted in 99.969.452 edges, the semi-automatic
  method used 131.637.777 edges, which was reduced to 30.000.577 edges with edge sparsification, The number of nodes was
  11.329.775.}.

\subsection{Accumulator based \mcH-arithmetic}

Constructing the task-graph for accumulator based arithmetic leads to similar results for the numerical tests. As is
shown in Figure~\ref{fig:timeseqaccu}, for sequential computations, the level-wise approach is again faster compared to
the semi-automatic method. Furthermore, the optimization from Section~\ref{sec:mergeaccudag} is much faster than the
combined approach, where a single DAG is constructed. Therefore, in the following, we will use the merged DAG by
default in all experiments.

\pgfdeclareimage[width=7.5cm]{timeseqaccu}{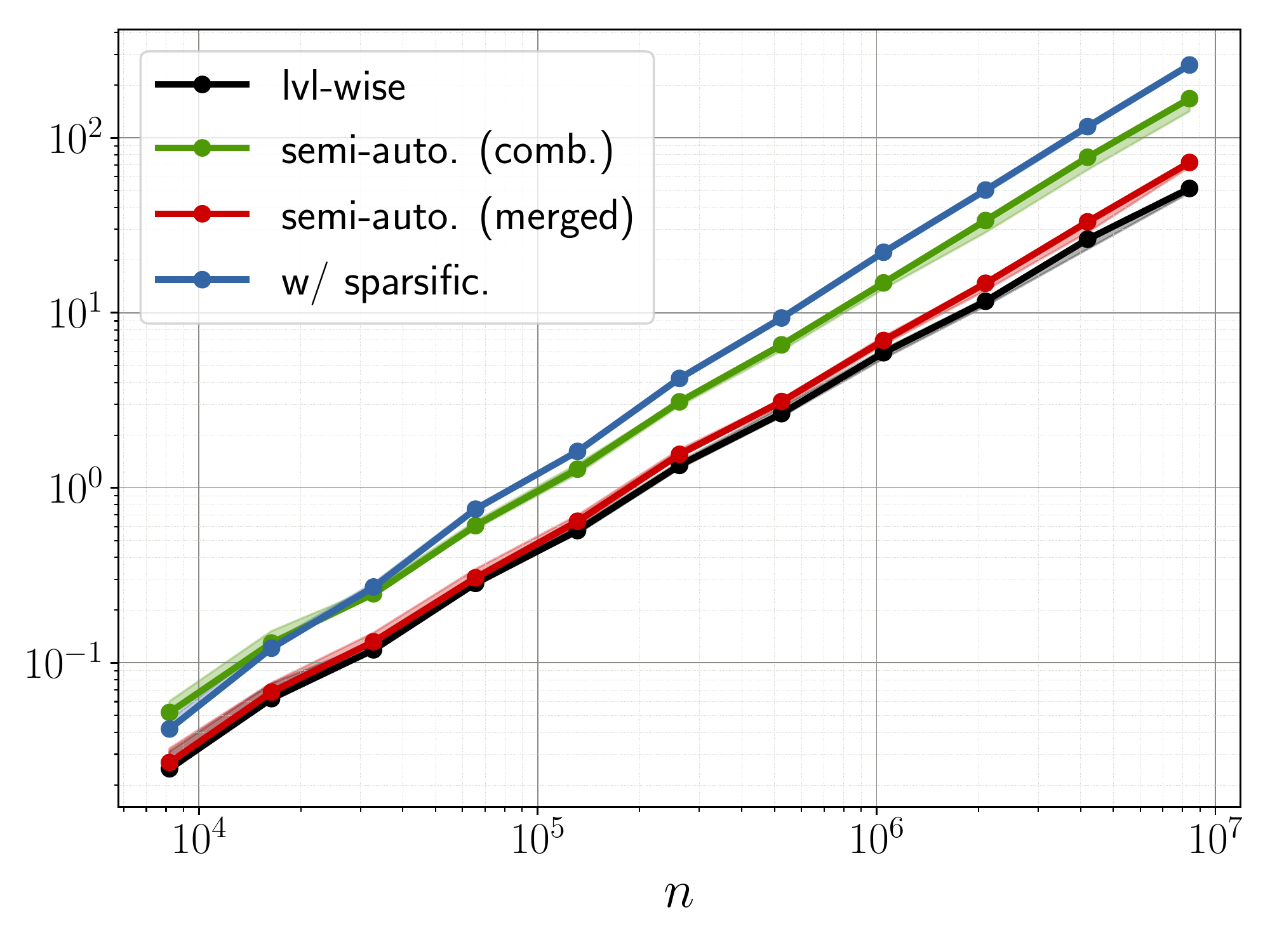}
\begin{figure*}[htb]
  \centering
  {%
  \begin{minipage}{0.45\linewidth}
    \pgfuseimage{timeseqaccu}
  \end{minipage}}
  \hspace{.1cm}
  {\begin{minipage}{0.45\linewidth}\small
    \begin{tabular}{rlll}
      \multicolumn{1}{c}{\(\mathbf{n}\)}
      & \multicolumn{1}{c}{\textbf{level-wise}}
      & \multicolumn{1}{c}{\textbf{semi-auto.}}
      & \multicolumn{1}{c}{\textbf{w/ sparsific.}} \\
      \toprule
      8.192     & \(2.29 \cdot 10^{-2}\) s & \(2.70 \cdot 10^{-2}\) s & \(4.19 \cdot 10^{-2}\) s \\
      16.384    & \(6.28 \cdot 10^{-2}\) s & \(6.86 \cdot 10^{-2}\) s & \(1.21 \cdot 10^{-1}\) s \\
      32.768    & \(1.20 \cdot 10^{-1}\) s & \(1.35 \cdot 10^{-1}\) s & \(2.71 \cdot 10^{-1}\) s \\
      65.536    & \(2.90 \cdot 10^{-1}\) s & \(3.16 \cdot 10^{-1}\) s & \(7.55 \cdot 10^{-1}\) s \\
      131.072   & \(5.64 \cdot 10^{-1}\) s & \(6.71 \cdot 10^{-1}\) s & \(1.61 \cdot 10^{0}\) s \\
      262.144   & \(1.36 \cdot 10^{0}\) s  & \(1.57 \cdot 10^{0}\) s  & \(4.22 \cdot 10^{0}\) s \\
      524.288   & \(2.67 \cdot 10^{0}\) s  & \(3.20 \cdot 10^{0}\) s  & \(9.34 \cdot 10^{0}\) s \\
      1.048.576 & \(5.90 \cdot 10^{0}\) s  & \(7.08 \cdot 10^{0}\) s  & \(2.22 \cdot 10^{1}\) s \\
      2.097.152 & \(1.18 \cdot 10^{1}\) s  & \(1.49 \cdot 10^{1}\) s  & \(5.03 \cdot 10^{1}\) s \\
      4.194.304 & \(2.64 \cdot 10^{1}\) s  & \(3.37 \cdot 10^{1}\) s  & \(1.16 \cdot 10^{2}\) s \\
      8.388.608 & \(5.21 \cdot 10^{1}\) s  & \(7.47 \cdot 10^{1}\) s  & \(2.61 \cdot 10^{2}\) s 
    \end{tabular}
    \vspace{.5cm}
  \end{minipage}}
  \caption{Sequential runtime of level-wise and semi-automatic task graph generation for accumulator based \mcH-arithmetic.}
  \label{fig:timeseqaccu}
\end{figure*}

Compared to the task graph without accumulators, the runtime is slightly faster on all cases, except for edge
sparsification. When comparing the number of nodes and edges in the corresponding DAGs, shown in
Table~\ref{tab:sizeaccu}, it can be seen that the number of nodes has increased due to \function{apply\_upd} and
\function{shift\_upd} tasks. However, the number of edges has decreased significantly. The reason is that accumulator
handling tasks now bundle update dependencies, which resulted in lots of unnecessary edges without edge
sparsification. For the same reason, benefit of edge sparsification is now smaller and with this the overhead of this
technique dominates, leading to a much higher runtime.

\pgfdeclareimage[width=7.5cm]{sizeaccu}{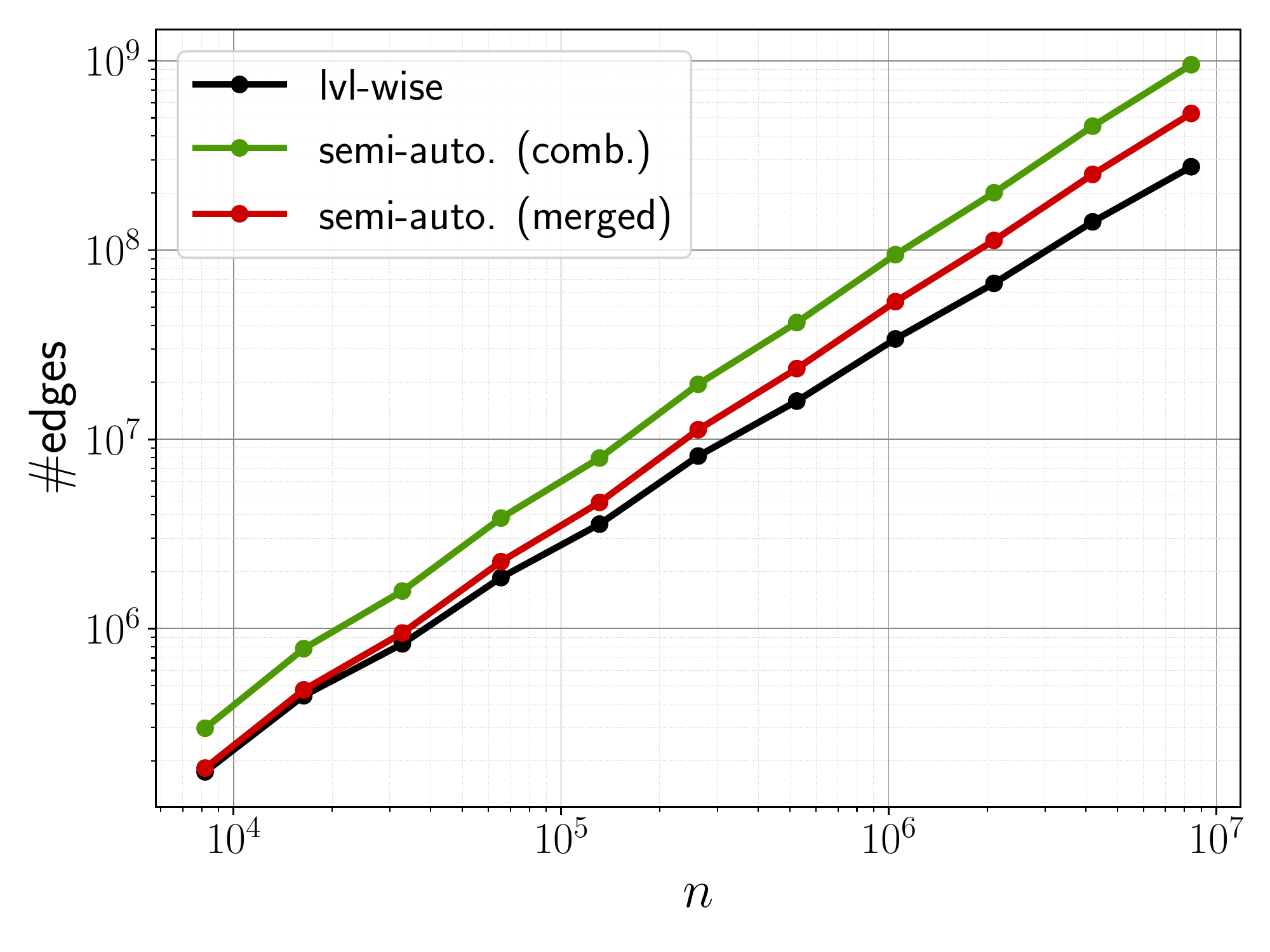}
\begin{table*}[htb]
  \centering
  \begin{tabular}{rrrrr}
    \multicolumn{1}{c}{\(\mathbf{n}\)}
    & \multicolumn{1}{c}{\textbf{\#nodes}}
    & \multicolumn{3}{c}{\textbf{\#edges}} \\
    &
    & \multicolumn{1}{c}{\textbf{level-wise}}
    & \multicolumn{1}{c}{\textbf{semi-auto.}}
    & \multicolumn{1}{c}{\textbf{w/ sparsific.}} \\
    \toprule
    8.192     &     45.027 &     174.941 &     183.218 &     183.840 \\
    16.384    &    101.584 &     441.522 &     474.277 &     479.169 \\
    32.768    &    177.121 &     829.991 &     947.430 &     963.174 \\
    65.536    &    378.372 &   1.858.564 &   2.256.501 &   2.282.233 \\
    131.072   &    708.698 &   3.561.068 &   4.640.641 &   4.563.089 \\
    262.144   &  1.583.958 &   8.157.734 &  11.238.839 &  10.738.999 \\
    524.288   &  3.019.337 &  15.919.483 &  23.627.396 &  21.654.670 \\
    1.048.576 &  6.325.151 &  33.913.026 &  53.345.529 &  46.811.203 \\
    2.097.152 & 12.220.179 &  66.666.176 & 112.490.323 &  94.020.901 \\
    4.194.304 & 25.543.336 & 140.856.257 & 250.940.470 & 200.095.310 \\
    8.388.608 & 49.346.180 & 275.508.834 & 527.100.659 & 398.294.409
  \end{tabular}
  \caption{Number of nodes and edges of the DAGs due to level-wise and semi-automatic task graph generation for accumulator based \mcH-arithmetic.}
  \label{tab:sizeaccu}
\end{table*}

When using multiple cores the parallel speedup without sparsification is reduced compared to standard
\mcH-arithmetic. The reason for this behaviour may be due to the reduced number of edges, which further reduces the
computational load per node in the graph. The opposite effect slightly increases the parallel speedup in the case
that edge sparsification is activated. Since now the number of edges is increased, the amount of work per node is
also slightly larger. In both cases, already for small problem sizes, the semi-automatic DAG generation is faster
compared to the level-wise approach.

\pgfdeclareimage[width=7.5cm]{timeparaccu}{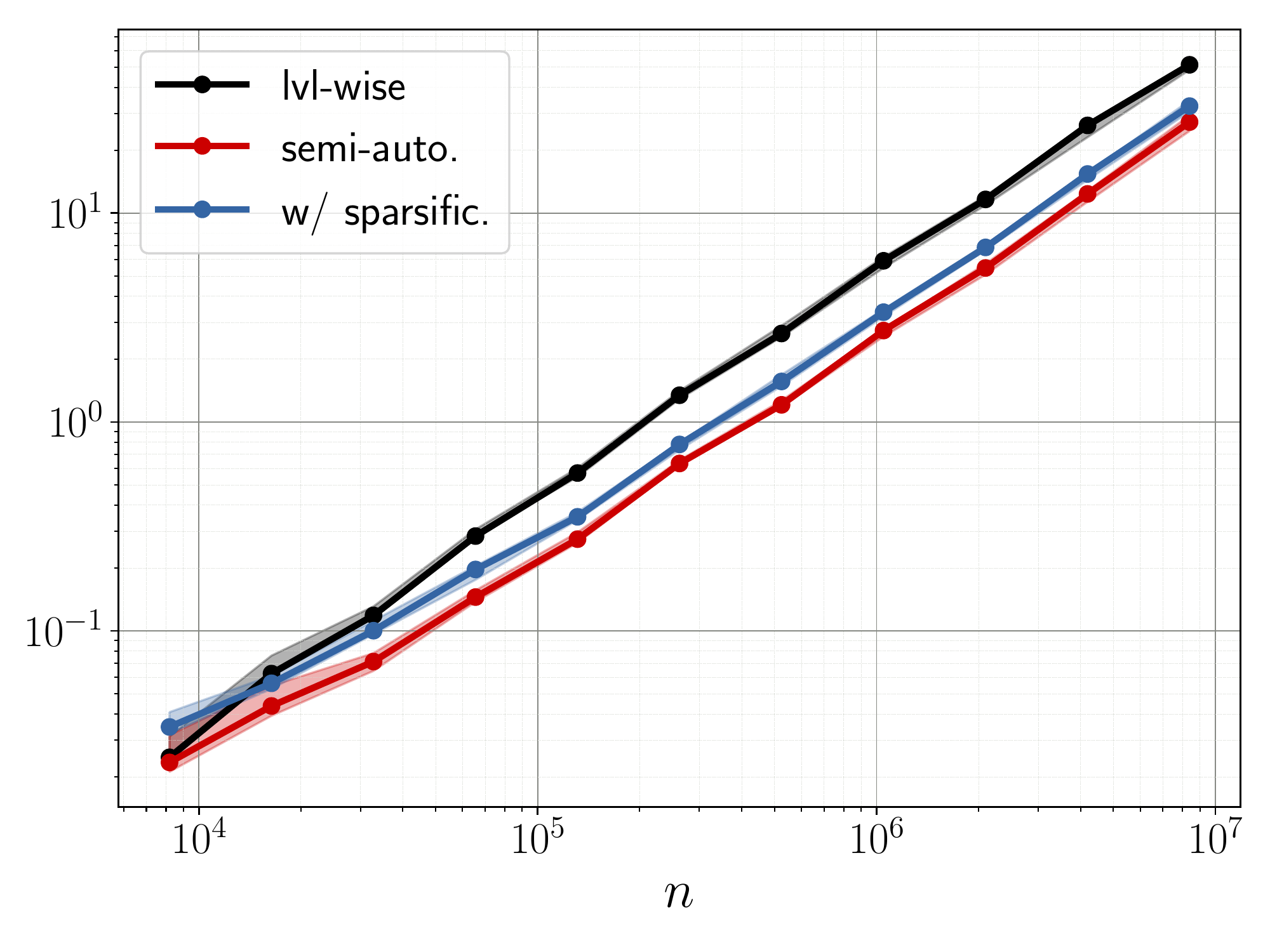}
\pgfdeclareimage[width=7.5cm]{speedupaccu}{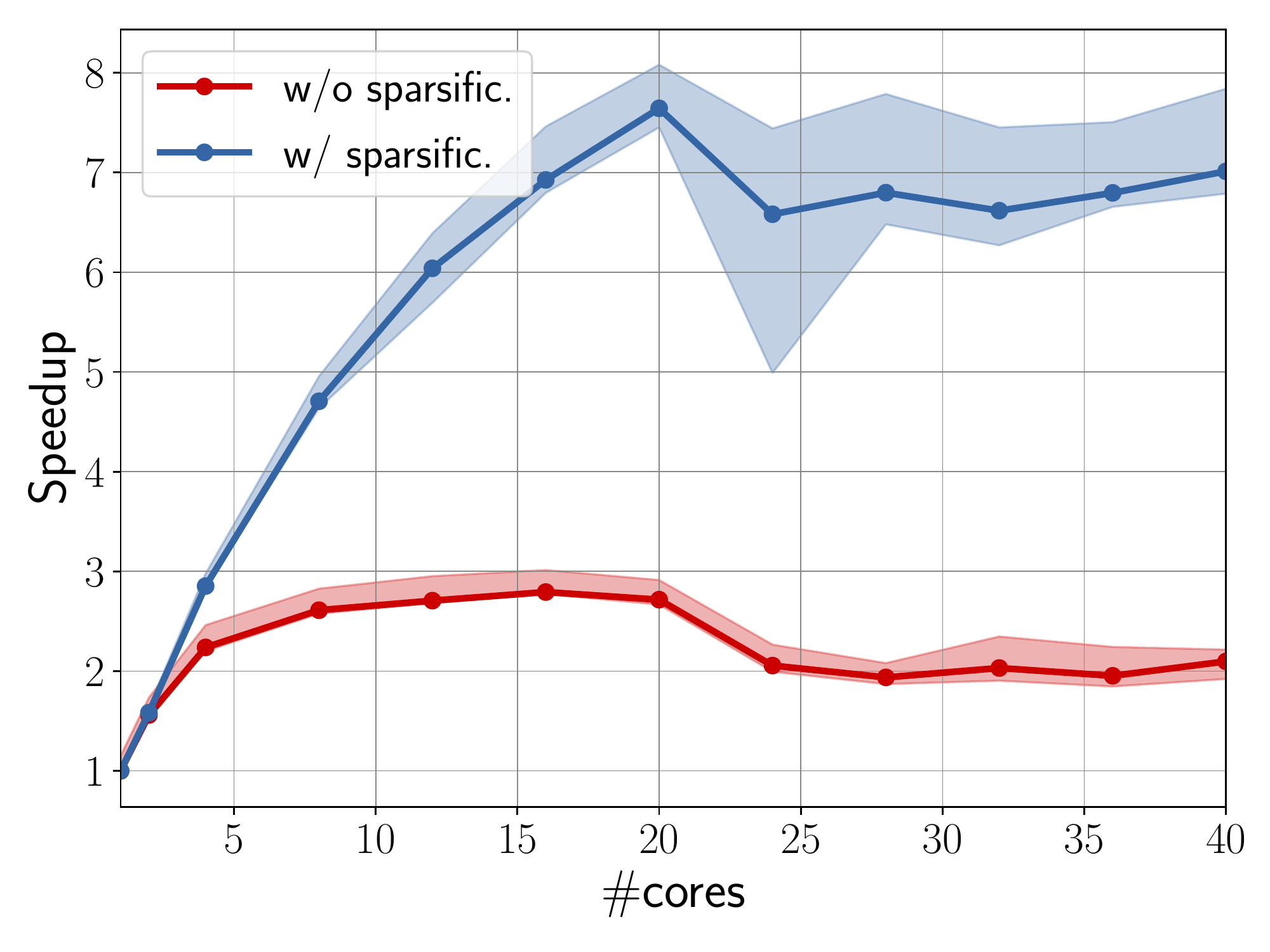}
\begin{figure*}[htb]
  \centering
  \begin{minipage}{0.45\linewidth}
    \pgfuseimage{timeparaccu}
  \end{minipage}
  \hspace{1cm}
  \begin{minipage}{0.45\linewidth}
    \pgfuseimage{speedupaccu}
  \end{minipage}
  \caption{Parallel runtime using one CPU (20 cores, left) and parallel speedup (right) of task graph generation for accumulator based \mcH-arithmetic.}
  \label{fig:speedupaccu}
\end{figure*}

When looking at the relative portion of the DAG construction on the full accumulator based \mcH-LU factorization, as
shown in Figure~\ref{fig:percdagaccu} for the Laplace SLP problem, again then semi-automatic approach is faster compared
to the level-wise method and the percentage is decreasing with larger problems. However, the task graph generation with
accumulators takes a significantly larger part in the full \mcH-LU factorization compared to standard \mcH-arithmetic.

\pgfdeclareimage[width=7.5cm]{percdagaccu}{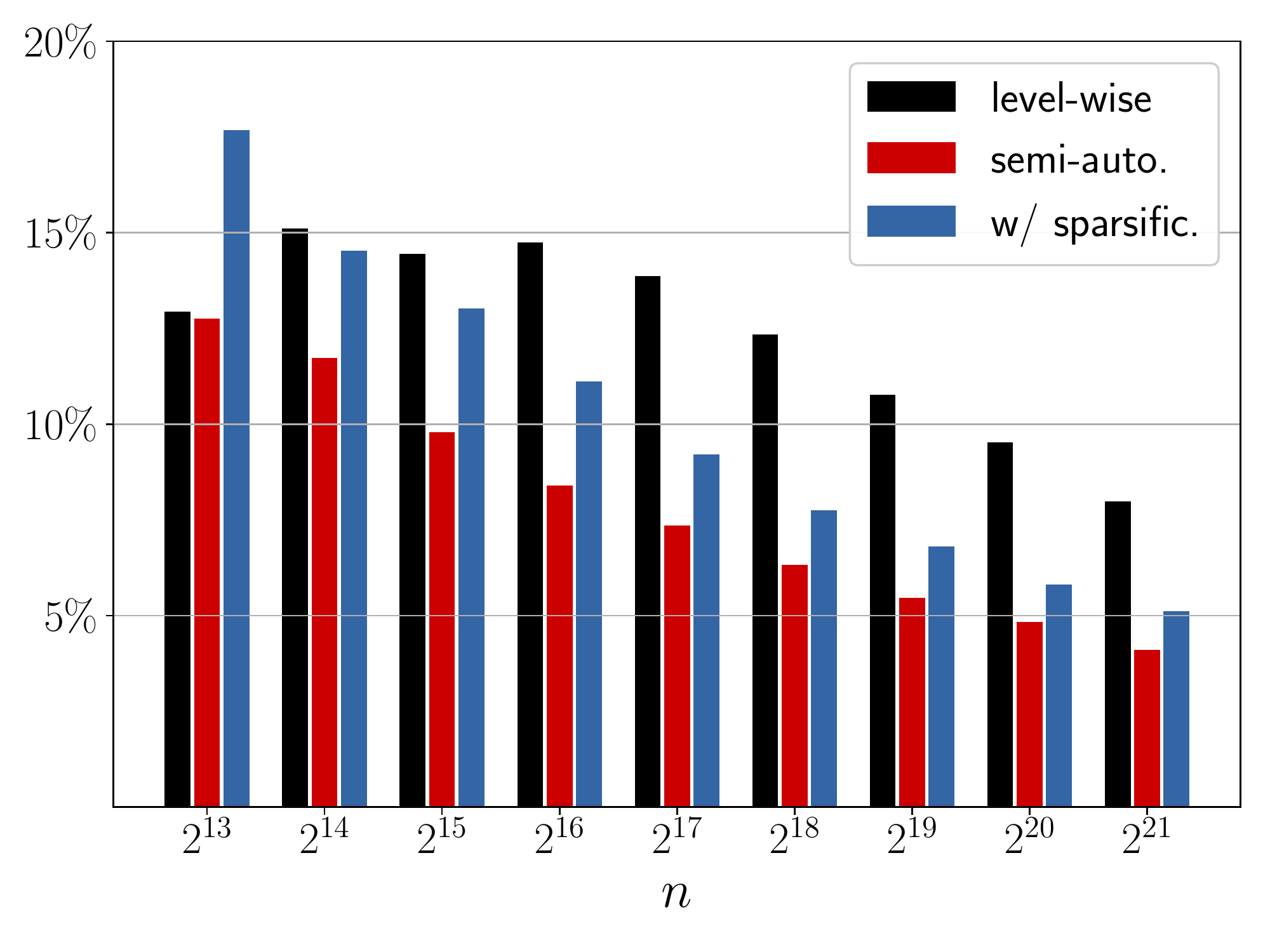}
\begin{figure}[htb]
  \centering
  \pgfuseimage{percdagaccu}
  \caption{Runtime percentage of task graph generation for entire accumulator based \mcH-LU algorithm.}
  \label{fig:percdagaccu}
\end{figure}

For the 1D model problem \eqref{eqn:1d} the task refinement overhead is even more dominant compared to standard
\mcH-arithmetic. This results in a higher runtime of semi-automatic task graph generation compared to the level-wise
method for all tested problem sizes as is shown in Figure~\ref{fig:accu1dsparse}~(left). 

In case of the sparse matrix example \eqref{eqn:pde}, the semi-automatic method is faster than the level-wise approach
for middle-sized problems due to less refinement overhead (Figure~\ref{fig:accu1dsparse}, right).

As for the Laplace SLP problem, also for the 1D model problem and the sparse matrix, edge sparsification does not result
in a lower runtime. Only memory consumption can be reduced.

\pgfdeclareimage[width=7.5cm]{timeaccu1d}{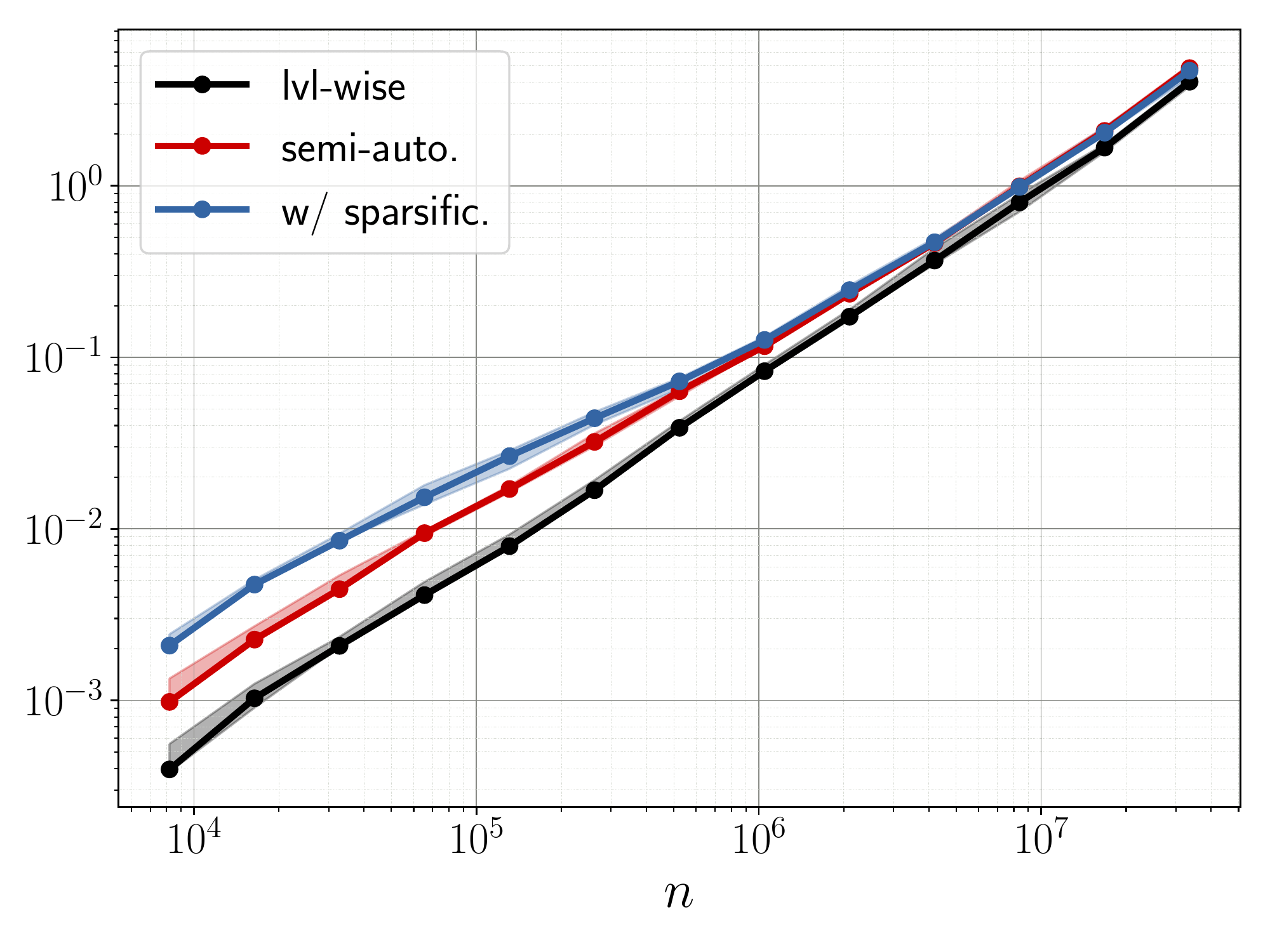}
\pgfdeclareimage[width=7.5cm]{timeaccusparse}{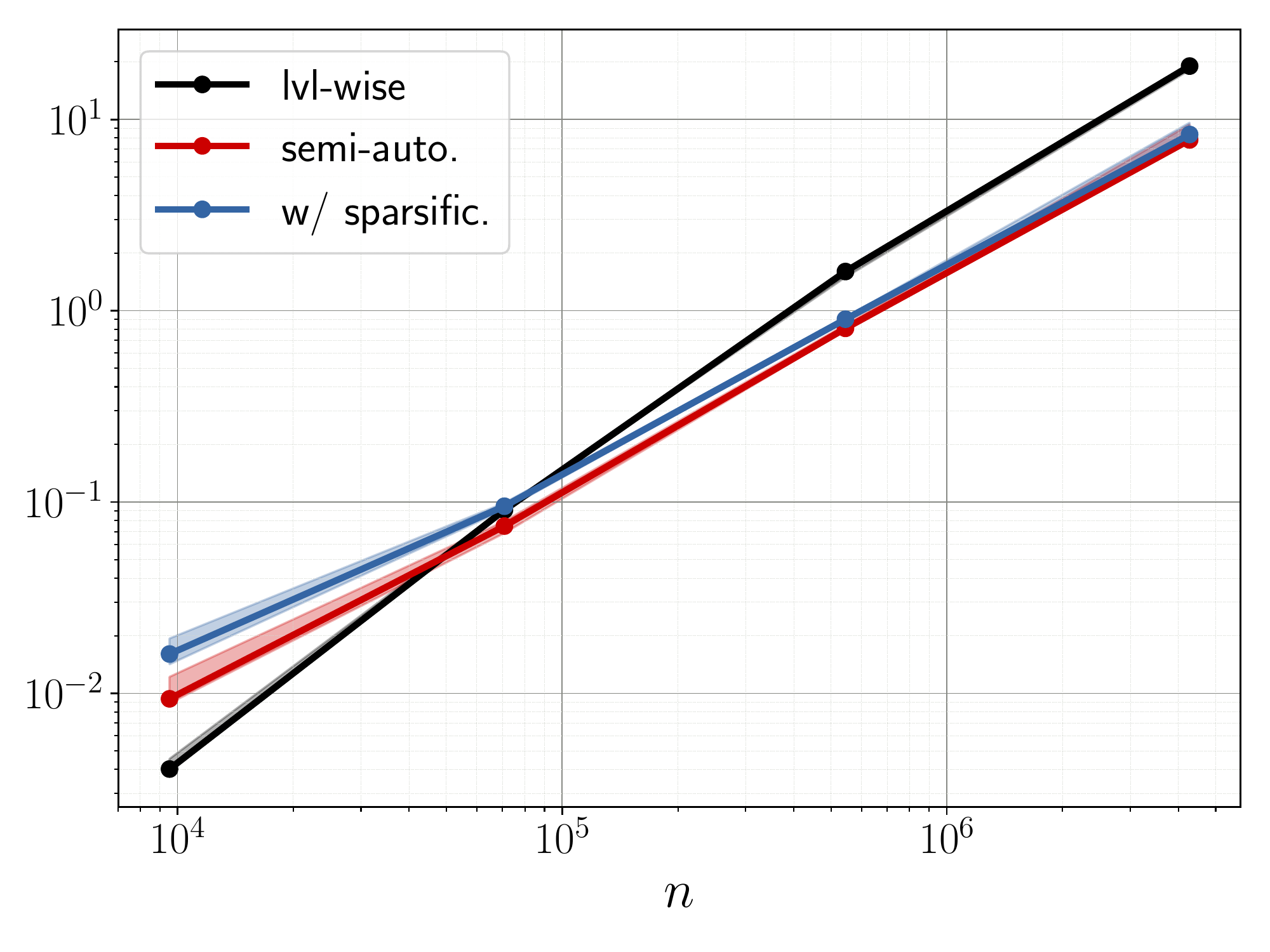}
\begin{figure*}[htb]
  \centering
  \pgfuseimage{timeaccu1d}
  \quad
  \pgfuseimage{timeaccusparse}
  \caption{Best runtime for task graph generation with accumulators for the 1D model problem (left) and the sparse matrix (right).}
  \label{fig:accu1dsparse}
\end{figure*}


\section{Conclusion} \label{sec:conclusion}

We have presented a new task graph generation procedure for \mcH-matrix arithmetic, which relies on the standard
recursive algorithms and the data dependencies expressed by the block index sets of the involved sub-blocks of the
\mcH-matrix. This significantly simplifies the implementation of task-based arithmetic for \mcH-matrices compared to
previous attempts while simultaneously keeping its high performance on many-core systems.

Accumulator based \mcH-matrix arithmetic fits naturally into the algorithm and shows excellent results on its own
compared to standard arithmetic.

Furthermore, since the new approach also permits parallelization, the task graph generation is also faster on multi-
and many-core CPUs compared to the previous, level-wise algorithm. However, the general parallel speedup is limited and
needs further investigation into how it can be improved. Though the DAG execution still takes much longer compared to
task graph generation, it also scales better with more CPU cores (see \cite{Kri:2013}). Therefore, for newer CPU
generations with even more CPU cores, a better parallel scaling behaviour of the task graph generation is needed to
maintain the current portion on the overall \mc-LU factorization procedure.

The next step is the application of the semi-automatic method on variations of the \mcH-matrix arithmetic, which were
previously not possible or extremely complicated. In fact, one such technique, currently in development and the topic of
an upcoming paper was the original motivation to investigate automatic task graph generation.

Another advantage of the semi-automatic approach, not discussed in this work, is the ability for DAG fusion, e.g., the
combination of separate task graphs for composed \mcH-arithmetic operations like \mcH-matrix inversion. This should
further increase the parallel efficiency of such operations on multi- and many-core systems.


\printbibliography

\end{document}